\documentclass[ reprint,
showpacs,  
aps,
pra,
floatfix,
]{revtex4-1}
\usepackage{amsmath,amsfonts,amsthm}
\usepackage{dsfont}
\usepackage{braket}
\usepackage{graphicx}
\usepackage{epstopdf}
\usepackage{dcolumn}
\usepackage{bm}
\usepackage{float}
\usepackage[caption=false]{subfig}
\usepackage{color}
\usepackage{float}
\usepackage{hyperref}

\begin{document}

\title{Classification of Quench Dynamical Behaviours in Spinor Condensates}
\author{Ceren B. Da\u{g}}
\email{cbdag@umich.edu}
\author{Sheng-Tao Wang}
\author{L.-M. Duan}
\affiliation{Department of Physics, University of Michigan, Ann Arbor, Michigan 48109, USA}
\date{\today}

\begin{abstract}
Thermalization of isolated quantum systems is a long-standing fundamental
problem where different mechanisms are proposed over time. We contribute to
this discussion by classifying the diverse quench dynamical behaviours of
spin-1 Bose-Einstein condensates, which includes well-defined quantum
collapse and revivals, thermalization, and certain special cases. These
special cases are either nonthermal equilibration with no revival but a
collapse even though the system has finite degrees of freedom or no
equilibration with no collapse and revival. Given that some integrable
systems are already shown to demonstrate the weak form of eigenstate
thermalization hypothesis (ETH), we determine the regions where ETH holds
and fails in this integrable isolated quantum system. The reason behind both
thermalizing and nonthermalizing behaviours in the same model under
different initial conditions is linked to the discussion of `rare'
nonthermal states existing in the spectrum. We also propose a method to
predict the collapse and revival time scales and find how they scale with the
number of particles in the condensate. We use a sudden quench to drive the
system to non-equilibrium and hence the theoretical predictions given in
this paper can be probed in experiments.

%\pacs{05.30.-d,05.70.−a,42.50.Gy}
\end{abstract}


% 34.50.-s	Scattering of atoms and molecules
% 67.85.De Dynamic properties of condensates; excitations, and superfluid flow
% 67.85.Fg	Multicomponent condensates; spinor condensates
% 05.30.Jp	Quantum statistical mechanics Boson systems
% 05.30.Rt	Quantum phase transitions

\maketitle

% \preprint{APS/123-QED}

% Force line breaks with\\

\affiliation{Department of Physics, University of Michigan, Ann Arbor,
Michigan 48109, USA} %Lines break automatically or can be forced with \\

%\author{}
%\affiliation{
%\affiliation{
%\affiliation{%


%\tableofcontents
%%%%%%%%%%%%%%%%%%%%%%%%%%%%%%%%%%%%%%%%%%%%%%%%%%%%%%%%%%%%%%%%%%%%%%%%%%%

\section{Introduction}

Understanding if and how isolated quantum systems driven out-of-equilibrium
thermalize has practical implications as well as being interesting from a
fundamental point of view. Being able to explain the thermalizing dynamics
in an isolated system is the key to have quantum thermal baths \cite{PhysRevLett.108.085303,PhysRevE.92.022104}.
Thermalization of quantum systems also sheds light on how the statistical
mechanics emerge from unitary dynamics of quantum mechanics \cite{Kaufman794,2016NatPh..12.1037N}. At the opposite
side, nonthermalizing quantum systems might be useful to store quantum
information in the protected degrees of freedom \cite{doi:10.1146/annurev-conmatphys-031214-014726,PhysRevB.88.014206}. 

Study of thermalization of
isolated quantum systems has a long history that starts with the development
of quantum mechanics itself \cite{vonNeumann2010} and can be understood in
the context of Eigenstate Thermalization Hypothesis (ETH) for isolated
systems \cite{PhysRevA.43.2046, PhysRevE.50.888, PhysRevLett.54.1879,
ETH,RevModPhys.83.863}. In this search to understand quantum thermalization, analogue concepts which are
important in the thermalization of classical systems have been drawn such as
the integrability of the system \cite{FPU, PhysRevLett.98.050405}. In this
paper, we study dynamics of the spin-$1$ spinor Bose-Einstein condensate
(BEC) system under single-mode approximation (SMA), which is known to be a
quantum-integrable model \cite{tabor1989chaos} based on its mean-field calculations \cite{PhysRevA.70.043614,PhysRevA.72.013602}. The consensus is that quantum-integrable systems do
not thermalize according to statistical ensembles, but they obey the
predictions of generalized Gibbs ensemble which takes into account the
conservation properties in the system Hamiltonian \cite%
{PhysRevLett.98.050405} in the aim of maximizing the entropy of the system
under study \cite{PhysRev.106.620}. However, it has also been shown that the
non-integrability does not always point to thermalization \cite%
{PhysRevLett.106.040401,PhysRevLett.98.210405,PhysRevLett.98.180601,PhysRevLett.119.030601}
and some integrable systems, e.g. Lieb-Liniger model and integrable spin
chains, do show thermalization in the form of weak ETH \cite%
{PhysRevLett.105.250401,PhysRevE.87.012125,PhysRevB.91.155123}. In fact, it
seems that what differentiates a quantum integrable system from a non-integrable
one in the context of thermalization is not that the system can thermalize
or not, but instead having `rare' nonthermal eigenstates in the spectrum
that do not disappear in the thermodynamic limit \cite%
{PhysRevLett.105.250401}. Our results of spinor condensate model support
this idea of quantum thermalization for a specific region of Hamiltonian
parameters, where we observe a spectrum composed of mostly `typical' thermal
states with some `rare' nonthermal ones. The exact diagonalization of spin-1
condensate under the SMA for realistic condensate sizes provides us the
opportunity to dig into the whole spectrum of eigenstates and determine the
regions where ETH is applicable based on the condition in Ref. \cite{ETH}.
Then we show that these regions in the spectrum are composed of `typical'
thermal eigenstates that lead to vanishing fluctuations and shrinking
support in the thermodynamic limit \cite{PhysRevLett.105.250401,
PhysRevE.87.012125}. We apply some other ETH indicators, as well, such as
the scaling of eigenstate expectation value differences \cite%
{PhysRevE.90.052105,PhysRevB.82.174411} and the scaling of the maximum
divergence from the microcanonical ensemble average \cite%
{PhysRevLett.119.030601} with the system size. The scaling exponents match
with each other and all of them point to the observation that in the
thermodynamic limit spinor condensates thermalize for certain initial
conditions (but not for all initial conditions), implying the weak form of
ETH. Given the fact that ultracold atoms provide a highly controllable and a
sufficiently isolated system \cite{RevModPhys.83.863}, we show that a spin-1
spinor condensate under the SMA could be a testbench to observe the
predictions of ETH for certain sudden quench parameters and the transition
between thermalization and nonthermalization without a need to add a
non-integrable perturbation to an integrable Hamiltonian \cite%
{PhysRevLett.106.025303, PhysRevLett.98.210405,PhysRevB.83.094431}. In fact,
being able to see this transition without breaking the integrability of the
model hints at that the thermalization is not directly tied to
non-integrability \cite{PhysRevLett.98.210405}. Instead, it might be more
relevant to consider the localization properties of the spectrum to observe
thermalizing behaviour in isolated quantum systems \cite%
{PhysRevB.83.094431,PhysRevB.82.174411}. Therefore, by invoking the analogy
between our model and the single quantum-particle hopping model and hence calculating
the participation ratios \cite{RevModPhys.80.1355} that is a widely-used
tool for Anderson models \cite{PhysRev.109.1492}, we show that the most localized
eigenstates in the spectrum (excluding the edges of the spectrum) are also the `rare' nonthermal eigenstates that cause nonthermalization behaviour in the system.

Quantum collapse and revivals are well-known phenomena observed in different
systems spanning from light-matter interactions in Jaynes-Cummings model 
\cite{scully1997quantum} to Bose-Hubbard models in optical lattices \cite%
{PhysRevLett.98.180601, will2010time-resolved} and the matter wave field of
a BEC \cite{greiner2002collapse}. This kind of behaviour is also expected in
discrete and finite systems due to the recurrence theorem \cite%
{PhysRev.107.337}. The possibility that spin-1 BEC under the SMA
might also demonstrate collapse and revivals has been suggested in Ref. \cite%
{PhysRevLett.81.5257,Pu200027} and a detailed analysis of collapses with
specific initial Fock states in this model has been given \cite%
{PhysRevA.60.1463}. These full-quantum model studies did not take the Zeeman
effects into account, partly because the model without Zeeman effects has
rotational symmetry and is analytically solvable via the introduction of
angular momentum-like operators in the Fock basis \cite{PhysRevLett.81.5257}%
. On the other hand, the experiments of the spinor BECs make use of the
quadratic Zeeman effect as a control parameter to sweep across the
well-established phase transitions \cite{RevModPhys.85.1191,
sadler2006spontaneous,PhysRevLett.111.180401} that spinor BECs have in their
mean-field representation \cite{NJP2003Zhang}. With the introduction of 
quadratic Zeeman effect, at the mean-field level the physics is mapped to an
analytical pendulum-like model \cite{PhysRevA.72.013602}. Some of the
mean-field predictions have been experimentally verified \cite%
{PhysRevA.90.023610}. However, the mean-field model cannot capture the
quantum collapse and revivals of the full-quantum Hamiltonian. In the second
part of our paper, we calculate the time scales for quantum collapse and
revivals in the spin-$1$ condensate model in the parameter region where they
exist and show that under realistic conditions and condensate sizes the
system equilibrates around its thermal value, validating the ETH for our
model. Finally, we discuss some particular parameter regions where we observe only equilibration but not
thermalization without quantum revivals in any time-scale of the
evolution. This is somewhat unexpected given the fact that our model is a
discrete system with finite degrees of freedom and the initial information
tends to recur in long-time scale for finite-size systems.
\section{Classification of Dynamical Behaviours under Sudden Quench of
Spin-1 Spinor Condensate}
\begin{figure}[tbp]
\centering
\subfloat[]{\label{fig:fig10a}\includegraphics[width=0.24%
\textwidth]{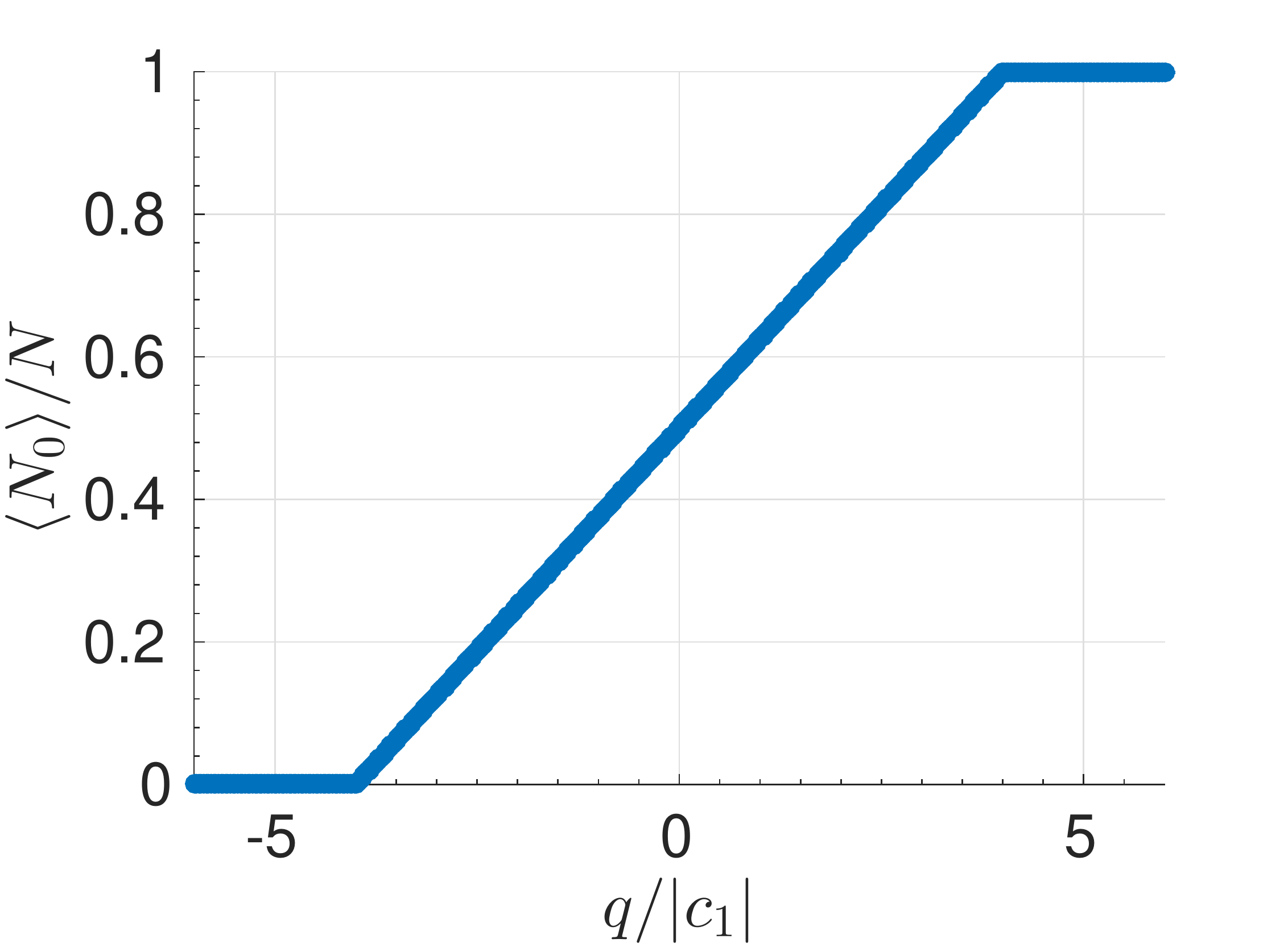}}\hfill \subfloat[]{\label{fig:fig10b}%
\includegraphics[width=0.24\textwidth]{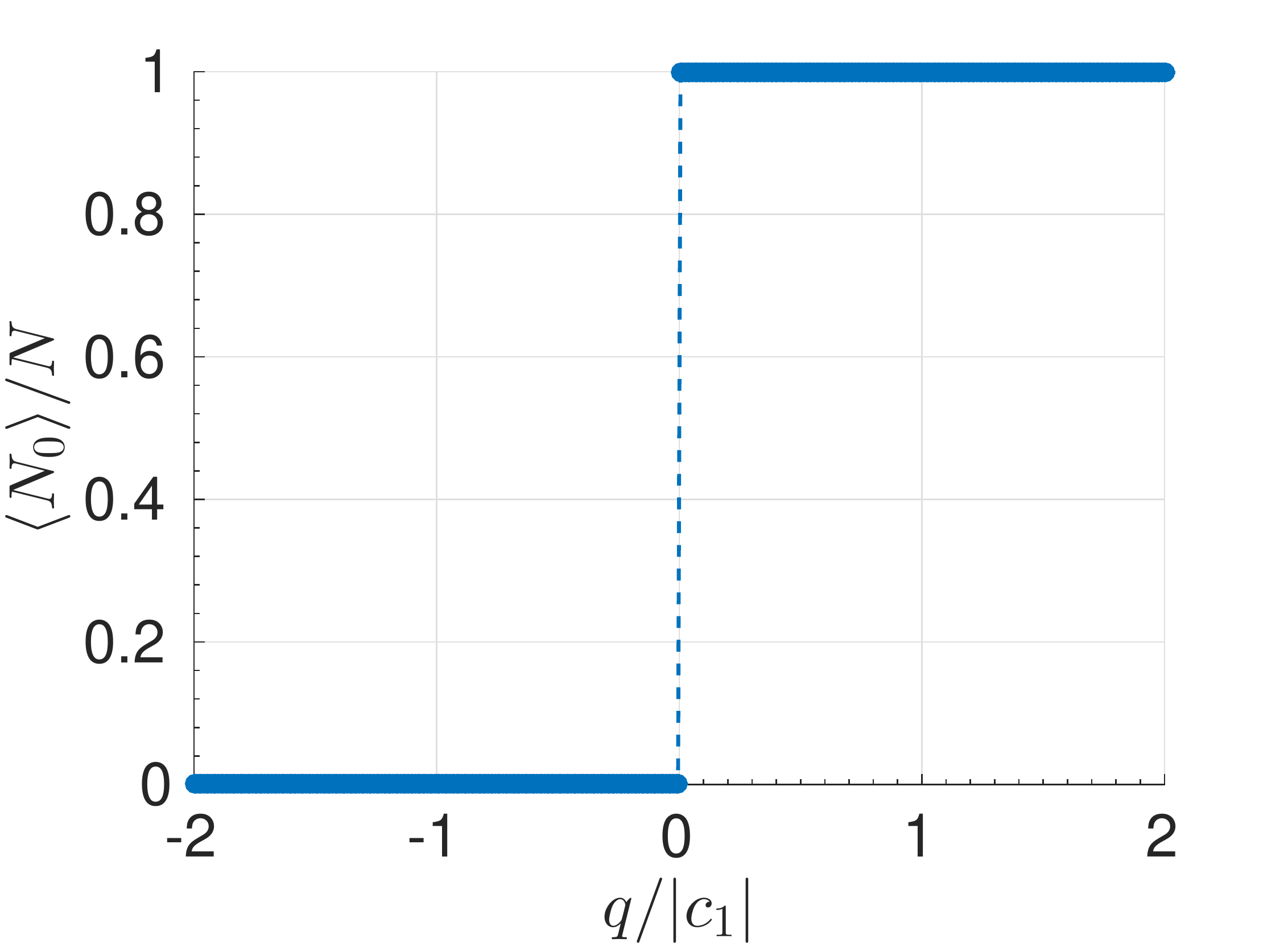}}
\caption{The ground state phase transitions for (a) ferromagnetic and (b)
anti-ferromagnetic interactions for $N = 10^4$ particles in the condensate
and zero total magnetization.}
\label{fig:fig10}
\end{figure}

The interaction Hamiltonian for a spin-$1$ BEC in the
second-quantization picture takes the form \cite{RevModPhys.85.1191} 
\begin{equation}
\hat{H}_{\text{int}}=\frac{1}{2}\int d\mathbf{r}\left( c_{0}^{\prime }:\hat{n%
}^{2}(\mathbf{r}):+c_{1}^{\prime }:\hat{F}^{2}(\mathbf{r}):\right) , \label{secondquantizationH}
\end{equation}%
where $::$ denotes the normal ordering. The coefficients in the interaction Hamiltonian depend on the scattering length and the atom mass through 
\begin{eqnarray}
c_{0}^{\prime } &=&\frac{4\pi \hbar ^{2}}{M}\frac{a_{0}+2a_{2}}{3},  \notag
\\
c_{1}^{\prime } &=&\frac{4\pi \hbar ^{2}}{M}\frac{a_{2}-a_{0}}{3}, 
\end{eqnarray}%
where $a_{0}$ and $a_{2}$ are the scattering lengths corresponding to a
total spin $0$ and a total spin $2$ of the colliding atoms. The operators in
the interaction Hamiltonian are defined by 
\begin{eqnarray}
\hat{n}(\mathbf{r}) &=&\sum_{m,n=-1}^{1}\hat{\psi}_{m}^{\dagger }(\mathbf{r}%
)I_{mn}\hat{\psi}_{n}(\mathbf{r}),  \notag \\
\hat{F_{\nu }}(\mathbf{r}) &=&\sum_{m,n=-1}^{1}\hat{\psi}_{m}^{\dagger }(\mathbf{r}%
)(F_{\nu })_{mn}\hat{\psi}_{n}(\mathbf{r}),
\end{eqnarray}%
where $\hat{\psi}_{m_{F}}$ $\left( \hat{\psi}_{m_{F}}^{\dagger }\right) $ is
the Bose field operator for the Zeeman state $m_{F}$. $I_{mn}$ and $%
(F_{\nu })_{mn}$ are the identity and spin-1 matrices, respectively and $\nu = x,y,z$ in the angular momentum operator $\hat{F_{\nu }}$. Also note that $\hat{F}^{2}(\mathbf{r}) = \hat{F}^{2}_x(\mathbf{r}) + \hat{F}^{2}_y(\mathbf{r}) + \hat{F}^{2}_z(\mathbf{r})$ in Eq. \eqref{secondquantizationH} and the identity matrix $I_{mn}$ results in the density operator $\hat{n}(\mathbf{r})$ for the condensate. For
sodium or rubidium alkali atoms, we have $|c_{0}^{\prime }|\gg
|c_{1}^{\prime }|$, so the symmetric part of the interaction Hamiltonian
dominates over the non-symmetric part. This observation leads to the
so-called single mode approximation (SMA), where we assume that the
condensate wave functions for each spin component $\phi _{m=-1,0,1}(\mathbf{r%
})$ are described by the same spatial wave function $\phi (\mathbf{r})$ as
in $\hat{\psi}_{m}\sim \hat{a}_{m}\phi (\mathbf{r}),\hspace{3mm}m=0,\pm 1$ \cite%
{RevModPhys.85.1191,PhysRevLett.81.5257,NJP2003Zhang,PhysRevA.66.011601}.
Then the spatial wave function $\phi (\mathbf{r})$ satisfies the
Gross-Pitaevskii equation which gives the spatial profile of our spin-1 Bose-Einstein condensate. With the normalization condition $\int d\mathbf{r}|\phi (\mathbf{r})|^{2}=1$, the interaction Hamiltonian reduces to
rotationally invariant $H_{\text{int}}=c_{1}\hat{L}^{2}/2N$, where $\hat{L}$ is the spin-1 angular momentum operator, $c_{1}=c_{1}^{\prime }N\int d\mathbf{r}|\phi (\mathbf{r})|^{4}$ and $N$ is
the total atom number, which has well-known analytical solutions \cite%
{PhysRevLett.81.5257}. In the experiment, an additional magnetic Zeeman
field is added to the system, which results in a competition between different
terms in the Hamiltonian and drives phase transitions \cite{PhysRevLett.111.180401}. The linear Zeeman term proportional to $\hat{L}_{z}=\hat{n}_{1}-\hat{n}_{-1}$ commutes with the other terms in the Hamiltonian, and
its effect is to conserve the magnetization. It has no influence on spin dynamics and
therefore can be dropped \cite{PhysRevLett.111.180401}. Adding the quadratic
Zeeman term, the Hamiltonian reduces to 
\begin{eqnarray}
H_{\text{int}} &=&c_{1}\frac{\hat{L}^{2}}{N}-q \hat{a}_{0}^{\dagger } \hat{a}_{0},  \notag \\
&=&\frac{c_{1}}{N}(\hat{a}_{1}^{\dagger } \hat{a}_{1}^{\dagger
}\hat{a}_{1}\hat{a}_{1}+\hat{a}_{-1}^{\dagger } \hat{a}_{-1}^{\dagger } \hat{a}_{-1} \hat{a}_{-1}-2 \hat{a}_{1}^{\dagger} \hat{a}_{-1}^{\dagger } \hat{a}_{1} \hat{a}_{-1}  \notag \\
&+&2 \hat{a}_{1}^{\dagger } \hat{a}_{0}^{\dagger } \hat{a}_{0}\hat{a}_{1}+2 \hat{a}_{-1}^{\dagger
} \hat{a}_{0}^{\dagger } \hat{a}_{0}\hat{a}_{-1}+2 \hat{a}_{0}^{\dagger } \hat{a}_{0}^{\dagger } \hat{a}_{1}\hat{a}_{-1}  \notag \\
&+&2\hat{a}_{1}^{\dagger }\hat{a}_{-1}^{\dagger }\hat{a}_{0}\hat{a}_{0})-q \hat{a}_{0}^{\dagger }\hat{a}_{0}.
\label{hamiltonian}
\end{eqnarray}%
Spin-1 BEC Hamiltonian with the quadratic Zeeman term gives rise to
different phases observed at the ground state due to the competition between
quadratic Zeeman effect and spin-mixing interaction \cite{PhysRevA.90.023610}. An adiabatic passage from one phase to another can create highly entangled states from product states as proposed in Ref.\ \cite%
{PhysRevLett.111.180401} and quite recently implemented in Ref.\ \cite%
{Luo620}. Fig.\ \ref{fig:fig10} shows the ground state quantum phase
transitions by observing the order parameter $\Braket{N_{0}}$, the number of
particles in Zeeman sublevel $\Ket{m=0}$, by varying the quadratic Zeeman
coefficient $q$.
In the rest of the paper, we study the dynamics of the system under a sudden
quench, i.e., we start from the ground state of the initial Hamiltonian $%
H_{i} $, which is $H_{\text{int}}$ (Eq.\ \eqref{hamiltonian}) with an
initial quadratic Zeeman term $q_{i}$, and abruptly quench the Zeeman field
to a final value $q_{f}$ with the final Hamiltonian denoted as $H_{f}$. The
dynamics and thermalization behaviour of the system are then investigated.
Both the sudden quench and the measurement of $\Braket{N_{0}}$, which is used as the
main observable in our study, can be readily performed in experiment \cite%
{PhysRevA.90.023610,PhysRevLett.102.125301,Hoang23082016,PhysRevLett.116.155301}%
.

\begin{figure}[tbp]
\centering
\subfloat[]{\label{fig:fig3a}\includegraphics[width=0.24%
\textwidth]{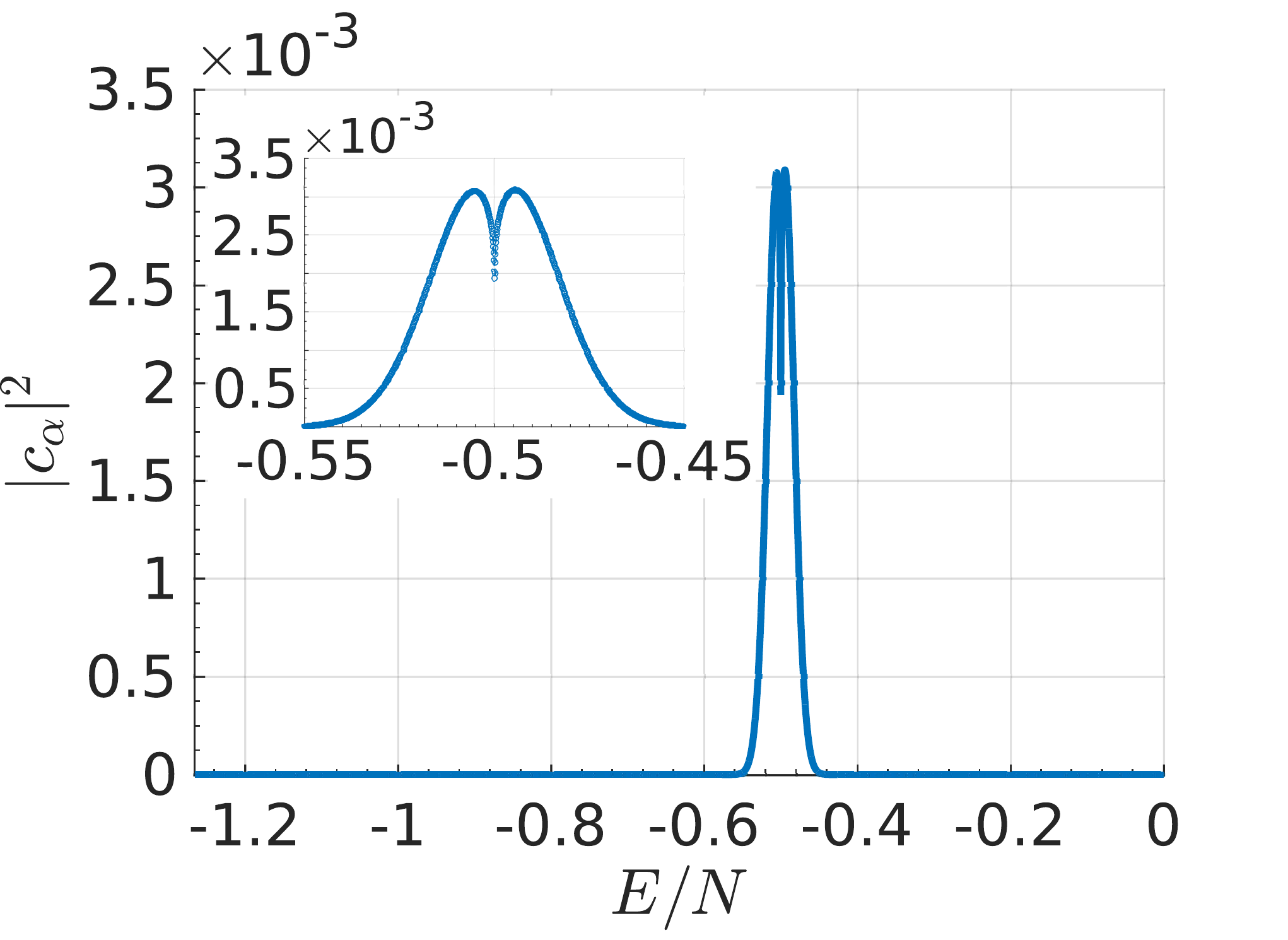}}\hfill \subfloat[]{\label{fig:fig3b}%
\includegraphics[width=0.24\textwidth]{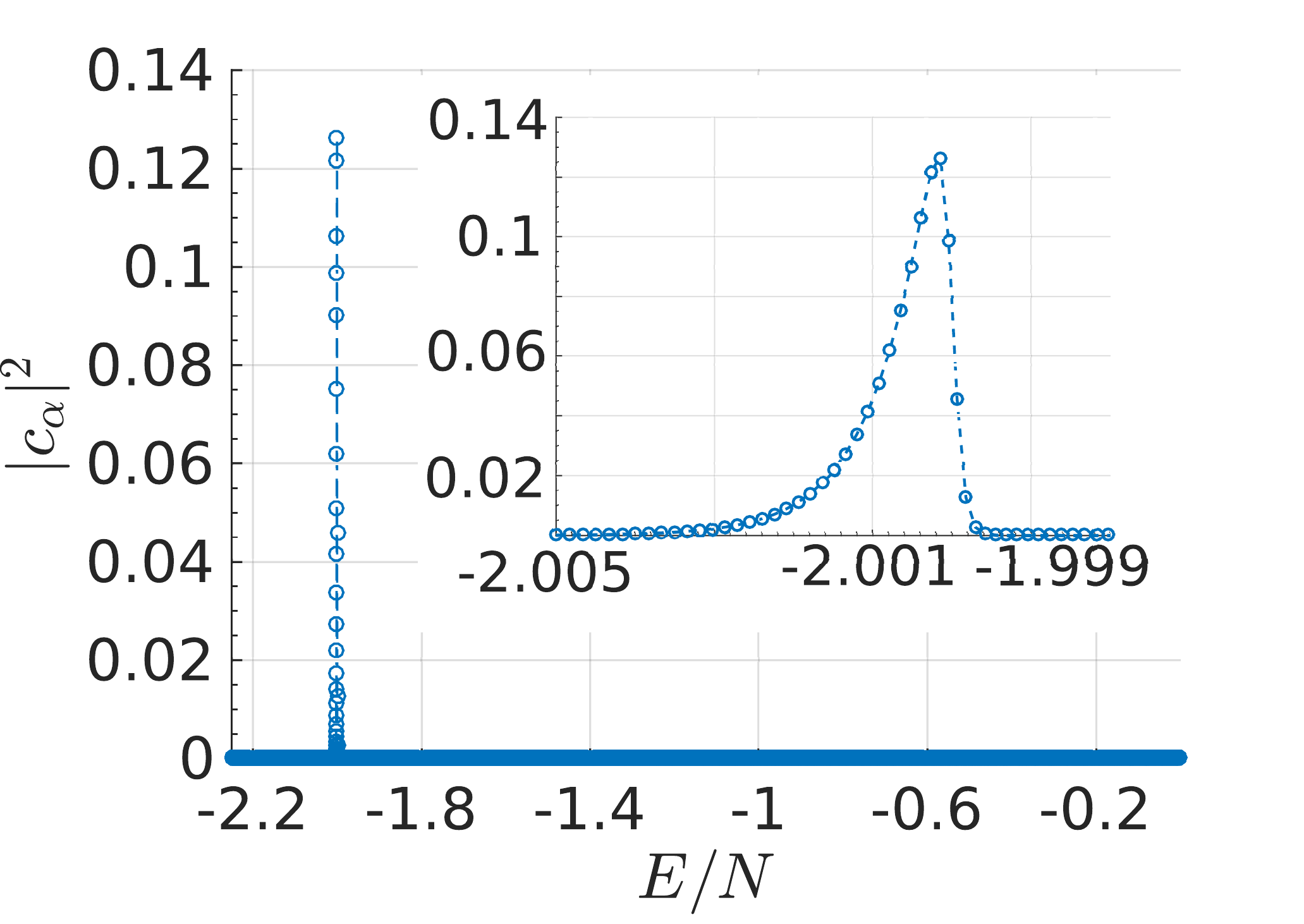}}
\par
\medskip \subfloat[]{\label{fig:fig3c}\includegraphics[width=0.23%
\textwidth]{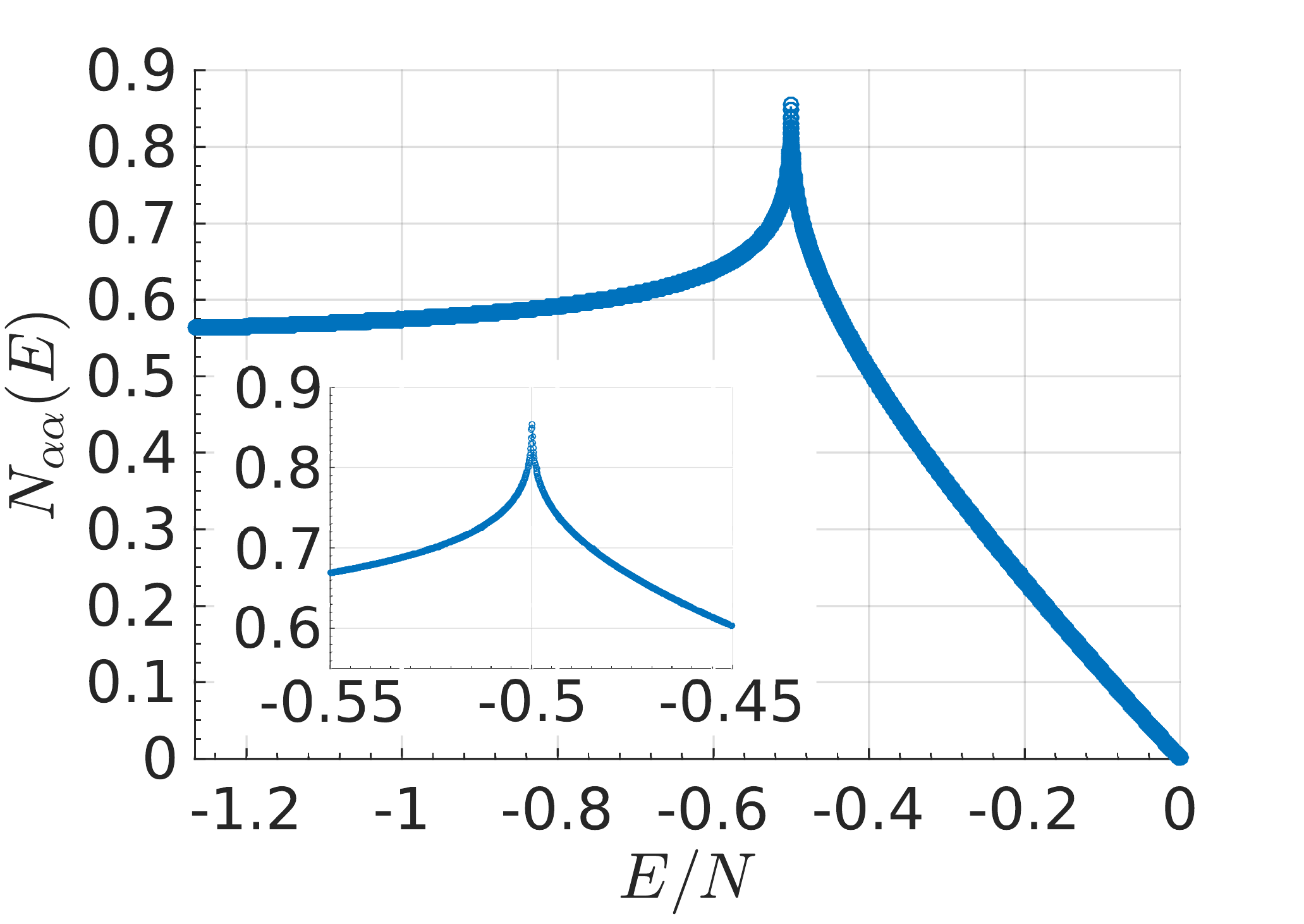}}\hfill 
\subfloat[]{\label{fig:fig3d}
\includegraphics[width=0.23\textwidth]{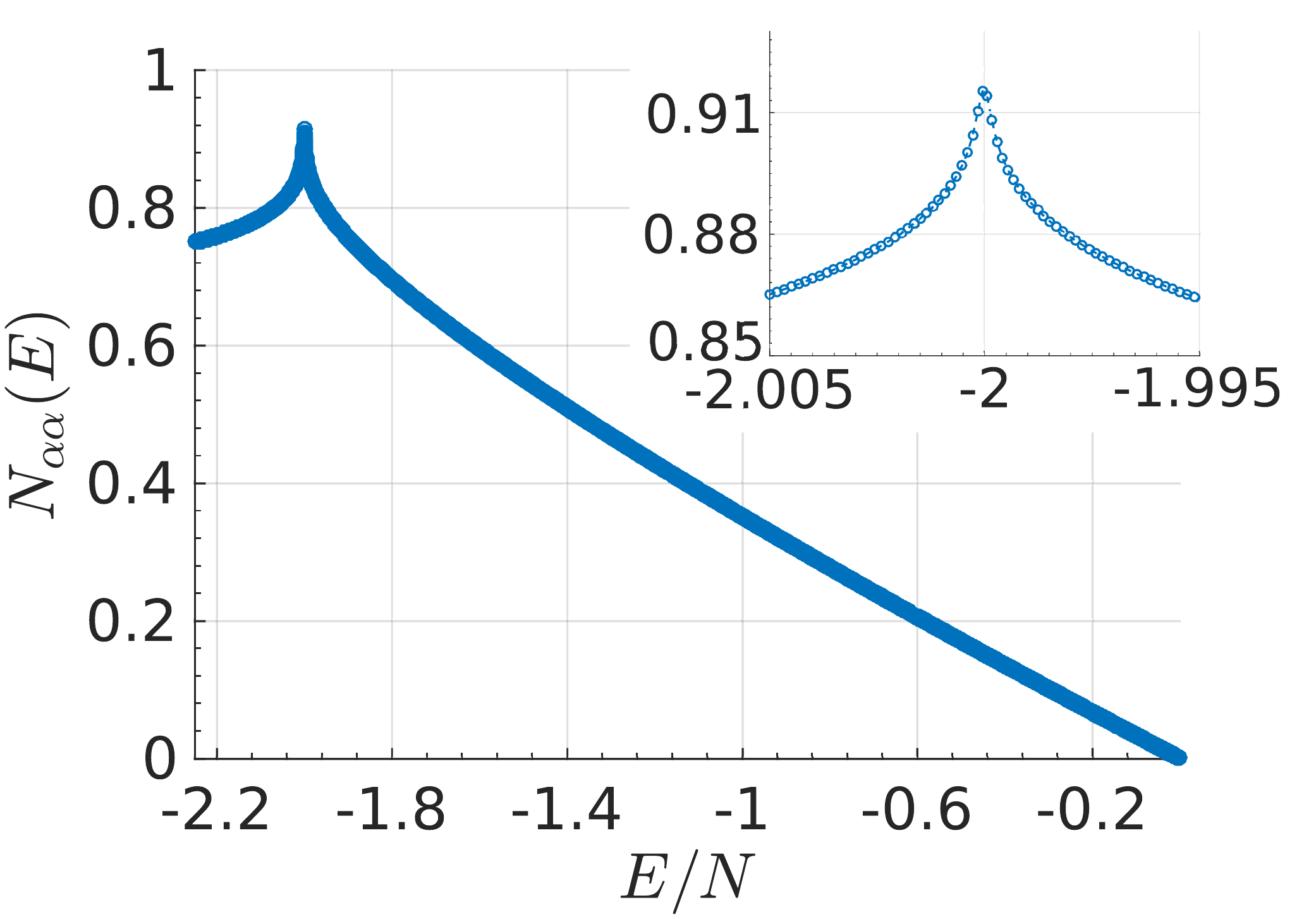}}
\caption{Eigenstate occupation numbers (EONs) $|c_{\protect\alpha }|^{2}$
for a ferromagnetic quench at (a) $q_{i}=-3$ to $q_{f}=0.5$ and (b) $q_{i}=4.1$ to $%
q_{f}=2 $ (focused on non-zero sections of the eigenspectrum in the insets) and their
corresponding eigenstate expectation values (EEVs) (focused on the nonlinear kink region in the insets) $N_{\protect\alpha 
\protect\alpha }$ at (c) and (d), respectively for a particle number of $10^{4}$ with respect to the energy density $E/N$.}
\label{fig:fig3}
\end{figure}

We now show how a dynamical phase transition (DPT) might be arising for spinor condensates via the sudden
quench based on an alternative definition of DPTs that takes the time-average of dynamical response as the order parameter \cite{2016arXiv160908482Z,2017arXiv170801044Z}. In our study, we start with the ground state, $\Ket{\psi(0)}$ of the
initial Hamiltonian $H_{i}$ with $q=q_{i}$. After a sudden quench of the
Zeeman coefficient $q$ to the value $q_{f}$, the initial state can be
expressed as 
\begin{equation}
\Ket{\psi(0)}=\sum_{\alpha }c_{\alpha }\Ket{\psi_{\alpha}},
\label{eqInitState}
\end{equation}%
where $\Ket{\psi_{\alpha}}$ are the eigenstates of the final Hamiltonian $%
H_{f}$. %so that the coefficients $c_{\alpha}$ can be calculated via
%\begin{eqnarray}
%c_{\alpha} &=& \Braket{\psi_{\alpha}|\psi(0)}.
%\end{eqnarray}
%Then the many-body wave function evolves as
%\begin{eqnarray}
%\Ket{\psi(t)} &=& e^{-i H_{\text{f}} t} \Ket{\psi(0)},\nonumber \\
%&=&  \sum_{\alpha} c_{\alpha} e^{-i E_{\alpha} t} \Ket{\psi_{\alpha}}.
%\end{eqnarray}
The number of atoms in the Zeeman sublevel $\Ket{m=0}$ can be written as 
\begin{eqnarray}
\Braket{N_0(t)} &=&\Bra{\psi(t)}N_{0}\Ket{\psi(t)},  \notag \\
&=&\sum_{\alpha ,\beta }c_{\alpha }^{\ast }c_{\beta }e^{-i(E_{\alpha
}-E_{\beta })t}N_{0,\alpha \beta },  \label{eq1}
\end{eqnarray}%
where $N_{0,\alpha \beta }=\Bra{\psi_{\alpha}}N_{0}\Ket{\psi_{\beta}}$ and $%
E_{\alpha }$ are the energy of the eigenstate $\Ket{\psi_{\alpha}}$ under
the final Hamiltonian $H_{f}$. The long-time average of $\Braket{N_0(t)}$
then should follow the diagonal ensemble prediction \cite{ETH, PhysRevLett.103.100403,
RevModPhys.83.863}, 
\begin{equation}
\overline{\Braket{N_0(t)}}_{t \rightarrow \infty}=\sum_{\alpha }|c_{\alpha }|^{2}N_{0,\alpha \alpha
},\label{predictionDE}
\end{equation}%
if the equilibration happens or when the phase coherence diminishes. In order to visualize this quantity, in
Figs.\ \ref{fig:fig3a} and \ref{fig:fig3b} we plot the eigenstate occupation
numbers (EONs) $|c_{\alpha }|^{2}$ for certain sudden quench parameters
(seen in the caption). EONs represent windows in the eigenspectrum where we
are allowed to peak into when we make a measurement. Figs.\ \ref{fig:fig3c}
and \ref{fig:fig3d} are plots of the corresponding eigenstate expectation
values (EEVs) $N_{0,\alpha \alpha }$. What we expect to see in the long-time
average of a sudden quench experiment is the summation of EEVs weighted with
EONs as shown by Eq.\ (\ref{predictionDE}). 
\begin{figure}[tbp]
\centerline{\includegraphics[width=0.5\textwidth]{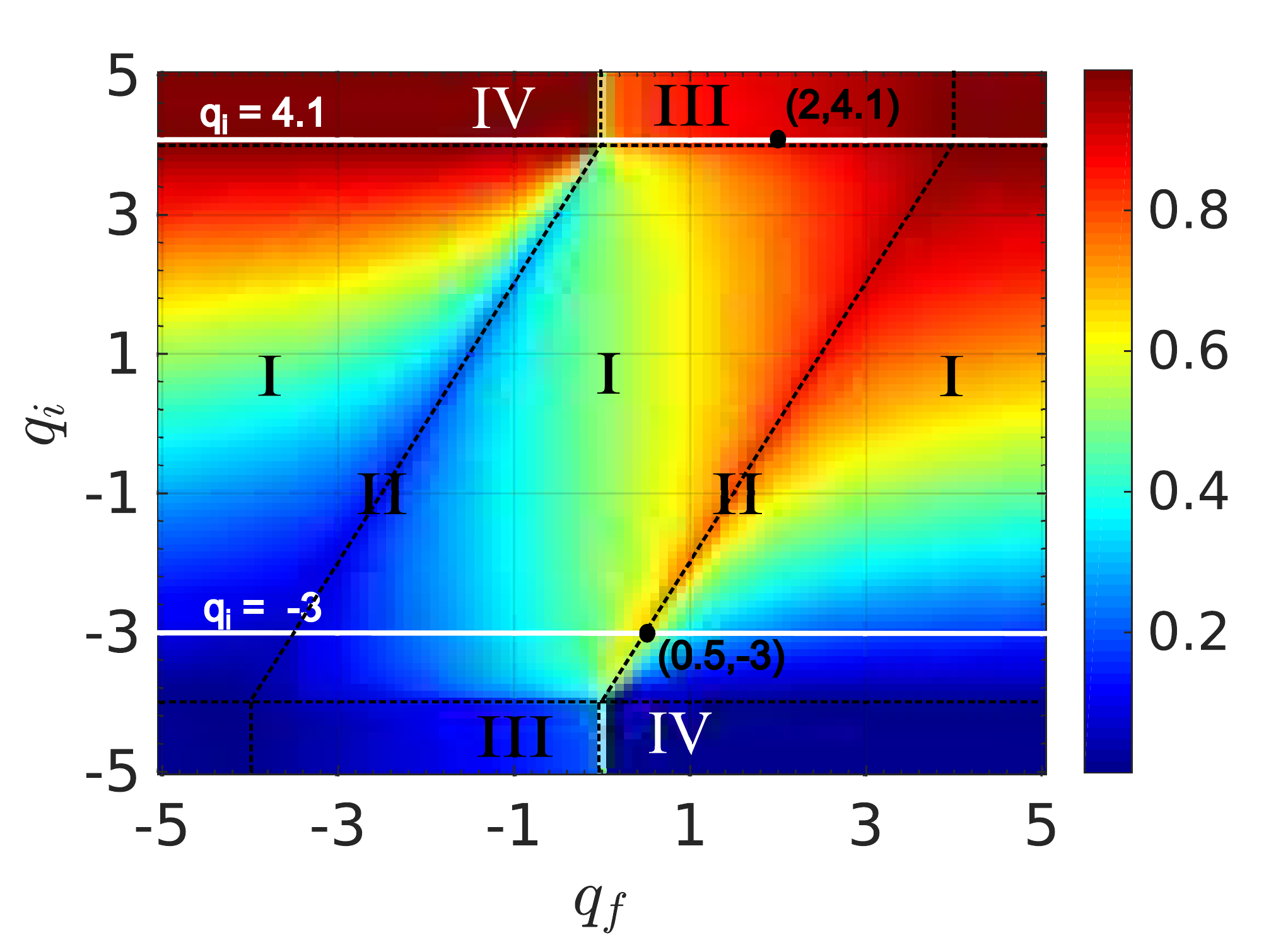}}
\caption{(Color online) The sudden quench map for the ferromagnetic case
with $q_{f}$ and $q_{i}$ on the x and y axes, respectively for $5\times
10^{3}$ particles. Color labels $\overline{\Braket{N_0(t)}}$ in the long
time limit.}
\label{fig:PD1}
\end{figure}
\begin{figure}[tbp]
\centerline{\includegraphics[width=0.5\textwidth]{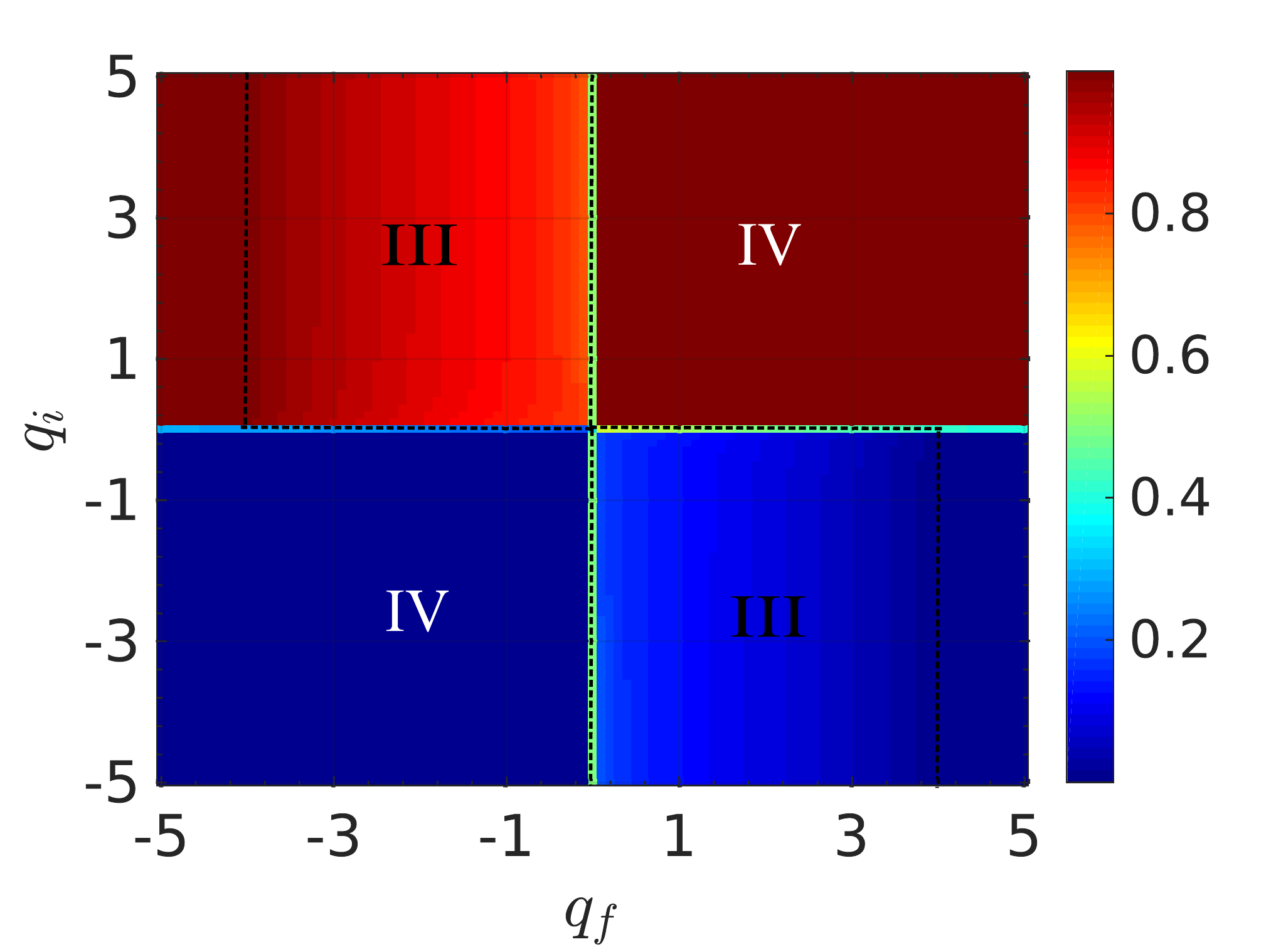}}
\caption{(Color online) The sudden quench map for the anti-ferromagnetic
case with $q_{f}$ and $q_{i}$ on the x and y axes, respectively for $5\times
10^{3}$ particles. Color labels $\overline{\Braket{N_0(t)}}$ in the long
time limit.}
\label{fig:PD2}
\end{figure}  

%\begin{widetext}
\begin{table*}[t]
\centering
\begin{tabular}{ccc}
Region & Boundaries & Dynamic Behaviour \\ \hline
I & $|q_i| < 4$, all $q_f$ except traces & ETH valid, well-defined collapse
and revivals \\ 
II & $|q_i| < 4$, traces of $q_f$ & nonthermal equilibration, collapse, no
revival \\ 
III & $q_i > 4$ $\&$ $0 < q_f < 4$ or $q_i < -4$ $\&$ $-4 < q_f < 0$ & no
equilibration, no collapse or revival \\ 
IV & the rest of the map & no non-equilibrium evolution \\ \hline
\end{tabular}%
\caption{The regions of the sudden quench map for the ferromagnetic case.}
\label{table_map}
\end{table*}
%\end{widetext}

Each point on sudden quench maps (Figs.\ \ref{fig:PD1} and \ref{fig:PD2})
corresponds to the prediction of diagonal ensemble (equilibration value if it happens, or
the time-average of the dynamic response of the system) when a sudden quench
is applied to the ground state from an initial Hamiltonian with $q_i$ to a
final Hamiltonian with $q_f$. Note that there are different regions on both
maps and the ferromagnetic sudden quench map is more diverse than the
anti-ferromagnetic one when the ground state is chosen as the initial state
of the non-equilibrium process. Due to the symmetry embedded in the
Hamiltonian for both interactions, one can obtain point symmetric version of
Fig.\ \ref{fig:PD1} (reflection with respect to the origin of the plot) with
anti-ferromagnetic interaction when the initial state is set as the
most-excited state of the Hamiltonian.

These maps capture the ground state phase transition points of both FM ($%
q=\pm 4$) and AFM ($q=0$) cases. In Fig.\ \ref{fig:PD2}, the upper half ($%
q_{i}>0$) of the map plane reveals two different regions with transition
points at $q_{f}=-4$ and $q_{f}=0$. Similarly for the lower half ($q_{i}<0$%
), we observe two regions with the transition points at $q_{f}=0$ and $%
q_{f}=4$. In Fig.\ \ref{fig:PD1}, for $|q_{i}|>4$ we see a similar behaviour
to Fig.\ \ref{fig:PD2} with transition points either at $q_{f}=0$ and $%
q_{f}=4$ (for $q_{i}>4$) or at $q_{f}=-4$ and $q_{f}=0$ (for $q_{i}<-4$). In
between $|q_{i}|<4$, the two transition points gradually shift as $q_{i}$
increases. In later sections, we are going to show that the sudden quench
maps also show us when we do and do not expect a thermal behaviour in our
system, similar to the non-equilibrium phase diagram given for Bose-Hubbard
model in Ref.\ \cite{PhysRevLett.98.180601}. Additionally it will provide us
a way to predict types of the dynamical behaviour in different time
scales. To give an idea of the regions on the maps, we summarized them in
Table \ref{table_map}. Although the non-equilibrium behaviour of these
regions will be explained in detail in the rest of the paper, we shortly
list them here. Region I is where the system equilibrates around its thermal
prediction after a collapse with a well-defined time-scale. It is also a
region where we observe clear quantum revivals due to finite-size
effects. Region II demonstrates nonthermal equilibration after a collapse,
but no clear collective-revival is observed for these points on the map. We
do not see equilibration, collapse or revival for the region III, instead we
observe an oscillatory behaviour around the system's PDE value due to the interference of a small number
of modes of the system. Finally in region IV, the initial state turns out to
be already in equilibrium with the quench Hamiltonian, giving us practically
a constant behaviour for all times.

\section{\label{sec:ETH} Eigenstate Thermalization Hypothesis in Spin-1 BEC}

When a system that is driven out-of-equilibrium equilibrates around a
thermal value predicted by a statistical ensemble, the process is called
thermalization. For isolated interacting bodies, microcanonical ensemble
describes the equilibrium predictions. In this context, ETH is a possible
pathway to thermalization and explains the match between the equilibration
value predicted by the diagonal ensemble after a quench (Eq. (\ref%
{predictionDE})) and the microcanonical thermal value \cite{ETH}. 
\begin{figure}[t]
\centerline{\includegraphics[width=0.5\textwidth]{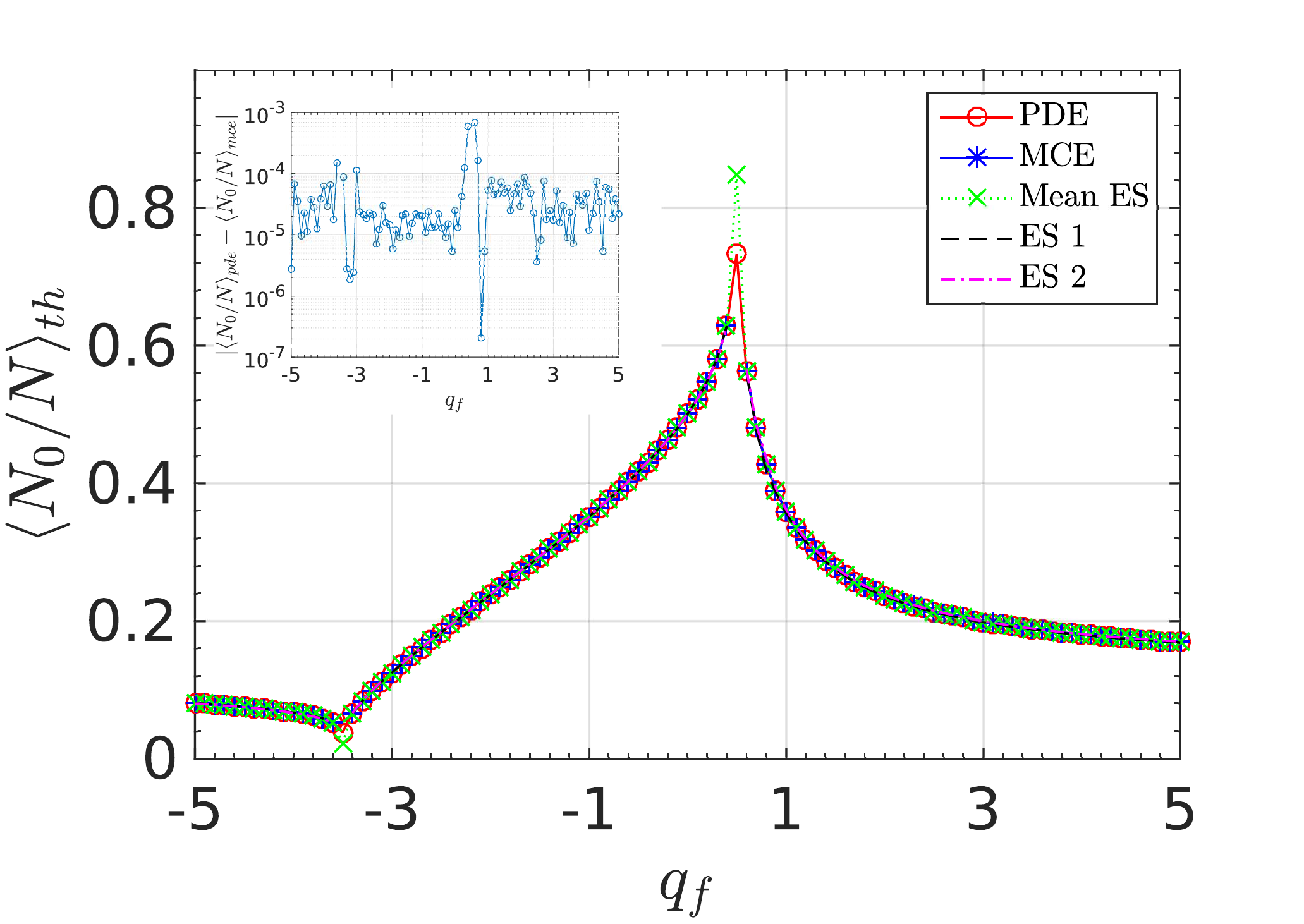}}
\caption{(Color online) The comparison of mean values predicted by diagonal
ensemble (PDE), microcanonical ensemble (MCE), the eigenstate corresponding
to the mean energy of the system (Mean ES) and arbitrary eigenstates in the
microcanonical energy window (ES 1 and ES 2), when the sudden quench is
applied from $q_{i}=-3$ to different $q_{f}$ values on the x-axis for ferromagnetic case. Each data
point is obtained with a simulation of $10^{4}$ particles. The inset shows
the difference between the diagonal and the microcanonical ensemble
predictions when it is possible to define a valid energy interval for the
microcanonical ensemble.}
\label{fig:thermalization1}
\end{figure}
%%ST: the inset label is too small, a little scrambled (q_{f}). The y-axis is a little weird in this way. maybe something like | [N_{0}]_{PDE} -  [N_{0}]_{MCE} |/N. And explain your inset and what you are trying to infer from there. The two dips at -3 and 0 may confuse readers a little bit too.  

\begin{figure}[t]
\centerline{\includegraphics[width=0.5\textwidth]{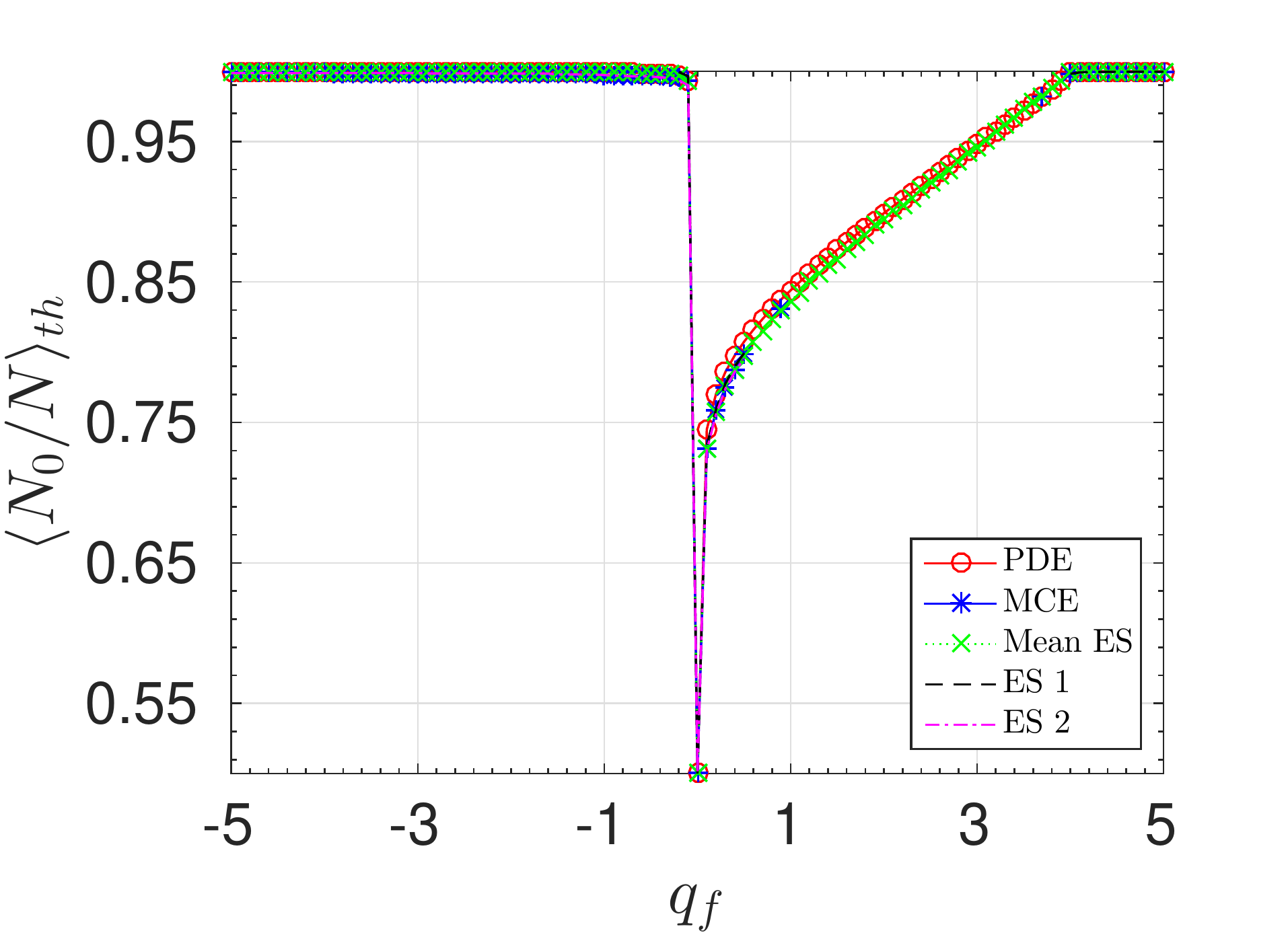}}
\caption{(Color online) The comparison of mean values predicted by diagonal
ensemble (PDE), microcanonical ensemble (MCE), the eigenstate corresponding
to the mean energy of the system (Mean ES) and arbitrary eigenstates in the
microcanonical energy window (ES 1 and ES 2), when the sudden quench is
applied from $q_{i}=4.1$ to different $q_{f}$ values on the x-axis for ferromagnetic case. Each
data point is obtained with a simulation of $10^{4}$ particles.}
\label{fig:thermalization2}
\end{figure}

The microcanonical ensemble is a statistical ensemble with a sufficiently
narrow energy interval that describes the equilibrium dynamics of an
isolated system \cite{kardar2007statistical}. In order to check the
prediction of microcanonical ensemble, we seek to define a narrow energy
window around the mean energy of the eigenspectrum. Refs.\ \cite%
{ETH,DunjkoBook,PhysRevLett.103.100403} emphasize the approximate linearity
of the EEVs in the microcanonical energy window in order to define a finite
and narrow energy window which will also ensure the validity of ETH. Based
on this idea, they state the following condition (which has been derived for the eigenstate thermalization to happen by Ref. \cite{0305-4470-29-4-003})
\begin{equation}
\left( \delta E\right) ^{2}|\Braket{N_{0}}^{\prime \prime }(E)/\Braket{N_{0}}(E)|\ll 1,\label{condition}
\end{equation}%
where $\delta E$ is the energy window, $\Braket{N_{0}} (E)$ is the EEV behaviour of the
system $N_{0,\alpha \alpha}$ as a function of the energy and $^{\prime
}$ denotes the differentiation with respect to energy. 
Another possibility implemented in Ref.\ \cite{ETH}
is to define the window based on a sensitivity analysis where the size of
the energy window chosen does not affect the thermal prediction of the
microcanonical ensemble (see App. A for a demonstration of this method for our model). We generate the finite and narrow microcanonical
energy windows for our model with a combination of these two ways. Figs.\ %
\ref{fig:thermalization1} and \ref{fig:thermalization2} show the regions
where the thermal prediction of diagonal ensemble (PDE) matches the
prediction of microcanonical ensemble (MCE), mean energy eigenstate (Mean
ES) and two arbitrary eigenstates (ES 1 and ES 2) in the microcanonical
energy window when it is possible to define one for a sudden quench from $%
q_{i}=-3$ and $q_{i}=4.1$ to various $q_{f}$ spanning from $-5$ to $5$,
respectively. It is important to note that the match happens only when the
EON window coincides with the approximately linear or constant parts
of the EEV plot. See Fig.\ \ref{fig:fig3} for the cases where the match does
not happen, so that the system fails to thermalize. Hence, we conclude that
the relaxation in the matching cases represents thermalization via ETH, when
we disregard the finite-size effects, e.g.\ a quantum
revival, which will be discussed in the next section.

\begin{figure}[tbp]
\centerline{\includegraphics[width=0.5%
\textwidth]{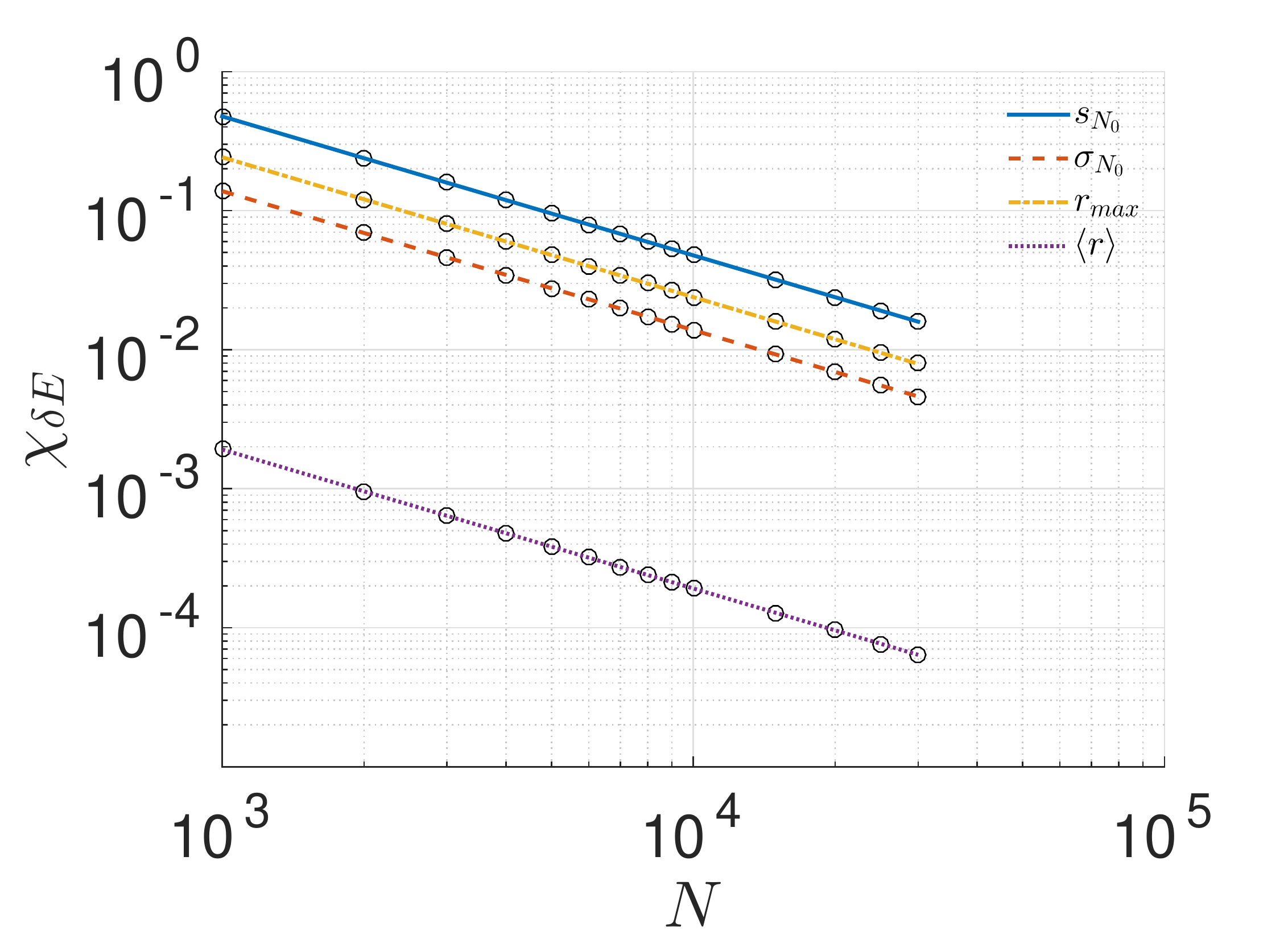}}\caption{(Color online) The system size scaling of the support $\chi_{\delta E} = s_{N_0}$ (solid-blue), the fluctuations (or the ETH noise) $\chi_{\delta E} = \sigma_{N_0}$ (dashed-red), the maximum divergence of EEV differences from the MC prediction $\chi_{\delta E} = r_{\text{max}}$ (dashed-dotted orange) and the average EEV difference $\chi_{\delta E} = \langle r_{n} \rangle_{\delta E}$ (dotted purple) for a
fixed energy interval when the interval is chosen right at the middle of the spectrum for $q_f = 3$. All of the scalings show a trend of $N^{-1}$ with $R^2 = 1$ where $R$ is the correlation coefficient.}
\label{fig:fig1AppB}
\end{figure}

In order to strengthen the argument that we see a nonthermal behaviour only
when EON captures the non-linear `kink' behaviour in the EEV spectrum, we
look at a couple of ETH indicators. These indicators are also used to
determine the form of ETH observed in the system, e.g. weak or strong, if
there is thermalization and they require an energy interval over the spectrum. It is possible
to define a microcanonical ensemble energy window at the linear region of
the spectrum with the methods mentioned above, while such a window is not
well-defined for the kink region. Since we want to compare two cases, we
define a fixed energy interval around the center of the spectrum. 
The first ETH indicator that we applied is the system size scaling of average EEV differences \cite%
{PhysRevB.82.174411,PhysRevE.90.052105}. An EEV difference is defined as
\begin{equation}
r_{n}=|\Bra{\psi_{n+1}}N_{0}\Ket{\psi_{n+1}}-\Bra{\psi_n}N_{0}\Ket{\psi_n}|,\label{Eq:EEVdiff}
\end{equation}%
for random eigenstate $\Ket{\psi_{n}}$ chosen in the energy interval and its adjacent state $\Ket{\psi_{n+1}}$. Regardless of
the interval size, when the interval encompasses the linear region as for $q_f
= 3$ in Fig. \ref{fig:fig1AppB}, we obtain the $N^{-1}$ scaling with $R^2 = 1$. Therefore the average of differences between EEVs vanish in the thermodynamic limit $N\rightarrow \infty $. Other indicators are: the ETH noise or fluctuations \cite{PhysRevLett.105.250401, PhysRevE.87.012125} 
\begin{eqnarray}
\sigma_{N_0} &=& \left( \frac{\sum_{\psi_n \in \delta E} \left[ \Bra{\psi_n} N_0 \Ket{\psi_n} - \Braket{N_0}_{mc,\delta E} \right]^2}{N_{\text{int}}} \right)^{1/2},
\label{ETHnoise}
\end{eqnarray}
where the $N_{\text{int}}$ is the number of eigenstates in the chosen interval, $\Braket{N_0}_{mc,\delta E}$ is the microcanonical prediction defined in the energy interval of $\delta E$ and $\Ket{\psi_n} \in \delta E $ are the eigenstates in the energy interval; the support of the eigenstate distribution in the energy interval \cite{PhysRevLett.105.250401, PhysRevE.87.012125}, 
\begin{eqnarray}
s_{N_0} &=& \text{max}_{\psi_n \in \delta E} \Bra{\psi_n} N_0 \Ket{\psi_n}  \notag \\
&-& \text{min}_{\psi_n \in \delta E} \Bra{\psi_n} N_0 \Ket{\psi_n},
\label{support}
\end{eqnarray}
 and the maximum divergence from the microcanonical ensemble prediction \cite%
{PhysRevLett.119.030601}, 
\begin{eqnarray}
r_{\text{max}} &=& \text{max}_n |\Bra{\psi_n} N_0 \Ket{\psi_n} - \Braket{N_0}_{\text{mc%
},\delta E}|,  \label{Eq:rmax}
\end{eqnarray}
in Fig. \ref{fig:fig1AppB} across the energy interval chosen. We obtain $N^{-1}$ scaling with $R^2 = 1$ for all these ETH indicators for the aforementioned case. The extracted scaling exponent of the support Eq. \eqref{support} clearly indicates in the thermodynamic limit all of the eigenstates in the energy interval contribute the same amount to the expectation value. Furthermore the rest of the ETH indicators, Eqs. \eqref{ETHnoise} and \eqref{Eq:rmax}, reveals that all of the EEVs in the energy interval converge to the microcanonical energy prediction $\Braket{N_0}_{\text{mc},\delta E}$ as $N \rightarrow \infty$. Also note that $N^{-1}$ scaling is not surprising, since the dimension of the Hilbert space is in the order of $N$ for our model. 
\begin{figure}[tbp]
\centerline{\includegraphics[width=0.5%
\textwidth]{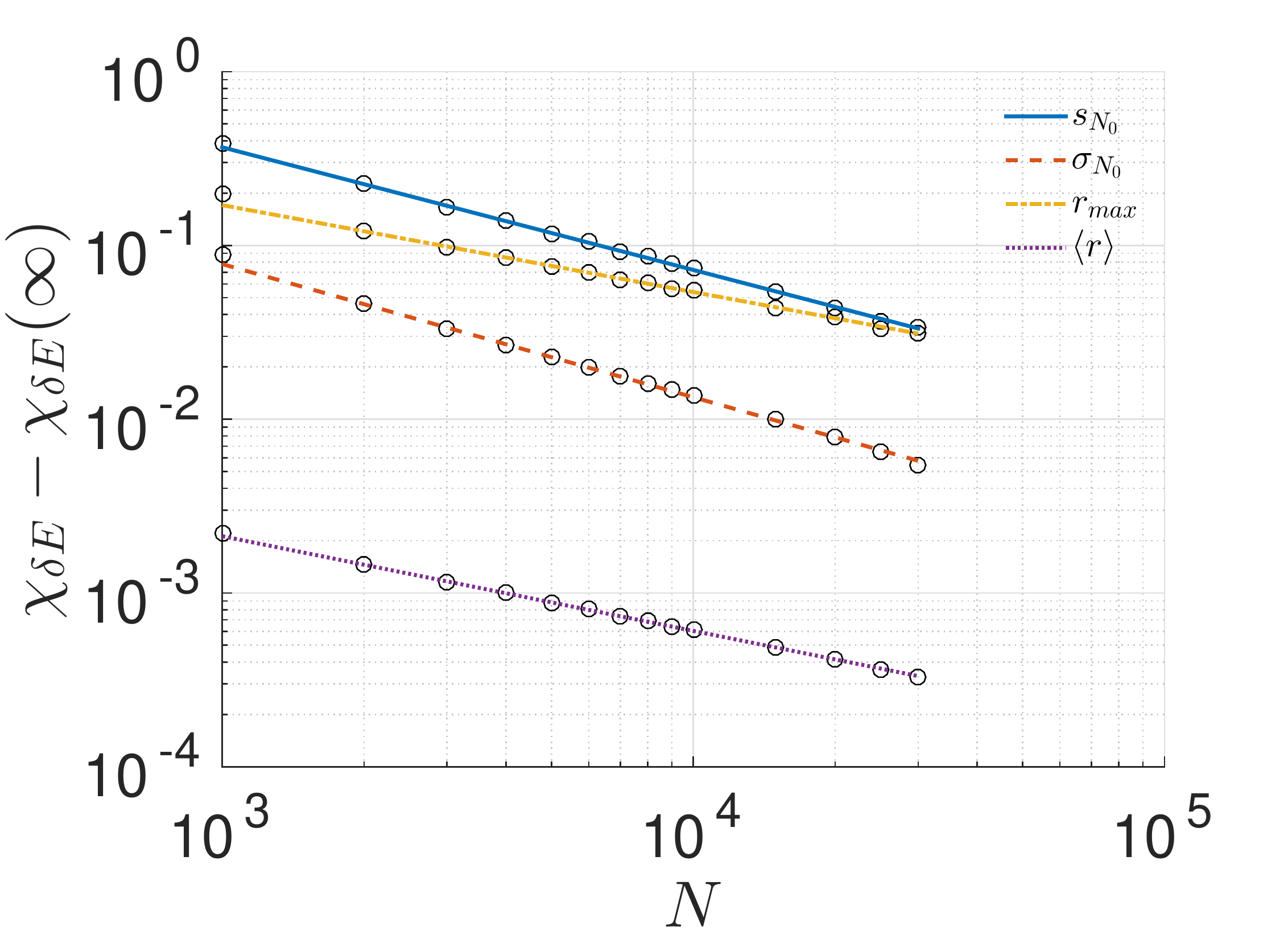}}
\caption{(Color online) The system size scaling of the support $\chi_{\delta E} = s_{N_0} = 0.07 + 48 N^{-0.71}$ (solid-blue) with $R^2 = 0.9997$, RMSE $=10^{-3}$, SSE $=10^{-5}$, the fluctuations (or the ETH noise) $\chi_{\delta E} = \sigma_{N_0} = 0.02 + 15.5N^{-0.77}$ (dashed-red) with $R^2 = 0.9994$, RMSE $=10^{-4}$, SSE $=10^{-6}$, the maximum divergence of EEV differences from the MC prediction $\chi_{\delta E} = r_{\text{max}} = 0.04 + 5.4N^{-0.5}$ (dashed-dotted orange)  with $R^2 = 0.998$, RMSE $=10^{-3}$, SSE $=10^{-5}$ and the average EEV difference $\chi_{\delta E} = \langle r_{n} \rangle_{\delta E} = 10^{-3} + 0.1 N^{-0.55}$ (dotted purple)  with $R^2 = 0.9996$, RMSE $=10^{-6}$, SSE $=10^{-10}$ for a
fixed energy interval when the interval is chosen right at the middle of the
spectrum for $q_f = 0.65$. Here $\chi_{\delta E}(\infty)$ stands for the offset value of the fitting. 
The RMSE and SSE stand for root mean square error and sum of squares of error, respectively.}
\label{fig:fig2AppB}
\end{figure}

The observation that all of the ETH indicators vanish in the thermodynamic limit for the linear regions of the
spectrum implies that ETH holds, even in the strong sense because of the
shrinking support \cite{PhysRevLett.105.250401}. However this is not the
case when the energy interval contains the kink region as seen in scaling plots for $q_f
= 0.65$ in Fig. \ref{fig:fig2AppB}. The scaling relation
for the support shows that the support still
exists in the thermodynamic limit when the kink region appears in the window. Therefore, we conclude that the kink
region is composed of nonthermal states that do not vanish in the
thermodynamic limit. Hence when the spectrum contains the kink region, the
whole spectrum will never have a shrinking support, violating the strong
form of ETH. Similarly, we observe a non-vanishing ETH noise when the kink
exists in the energy interval (dashed line in Fig. \ref{fig:fig2AppB}). In literature,
the fluctuations are expected to vanish away in the thermodynamic limit for
the weak form of ETH to hold \cite{PhysRevLett.105.250401}. However, we see
that they do not disappear when the interval includes the kink eigenstates.
This matches with the fact that we do not see thermalization when the
initial state overlaps with the kink eigenstates. Therefore, we can clearly
conclude that the kink eigenstates are nonthermal states that cause
nonthermalization when the initial state is chosen carefully to overlap with
them (Regions II and III on sudden quench maps). As a result, we argue that when the kink region exists in the spectrum
($|q| < 4$) not all initial states can lead the system to thermalization
even in thermodynamic limit. However due to the rarity of these nonthermal
states, most of the initial states will result in thermalization (Region 1
on sudden quench map Fig. (\ref{fig:PD1}). Therefore, the weak form of ETH
holds for $|q| < 4$, and otherwise ETH holds in the strong sense (based on
the shrinking support for all spectrum) since kink region disappears when we choose $|q| > 4$.
\begin{figure}[tbp]
\centering
\subfloat[]{\label{fig:fig4a}\includegraphics[width=0.24%
\textwidth]{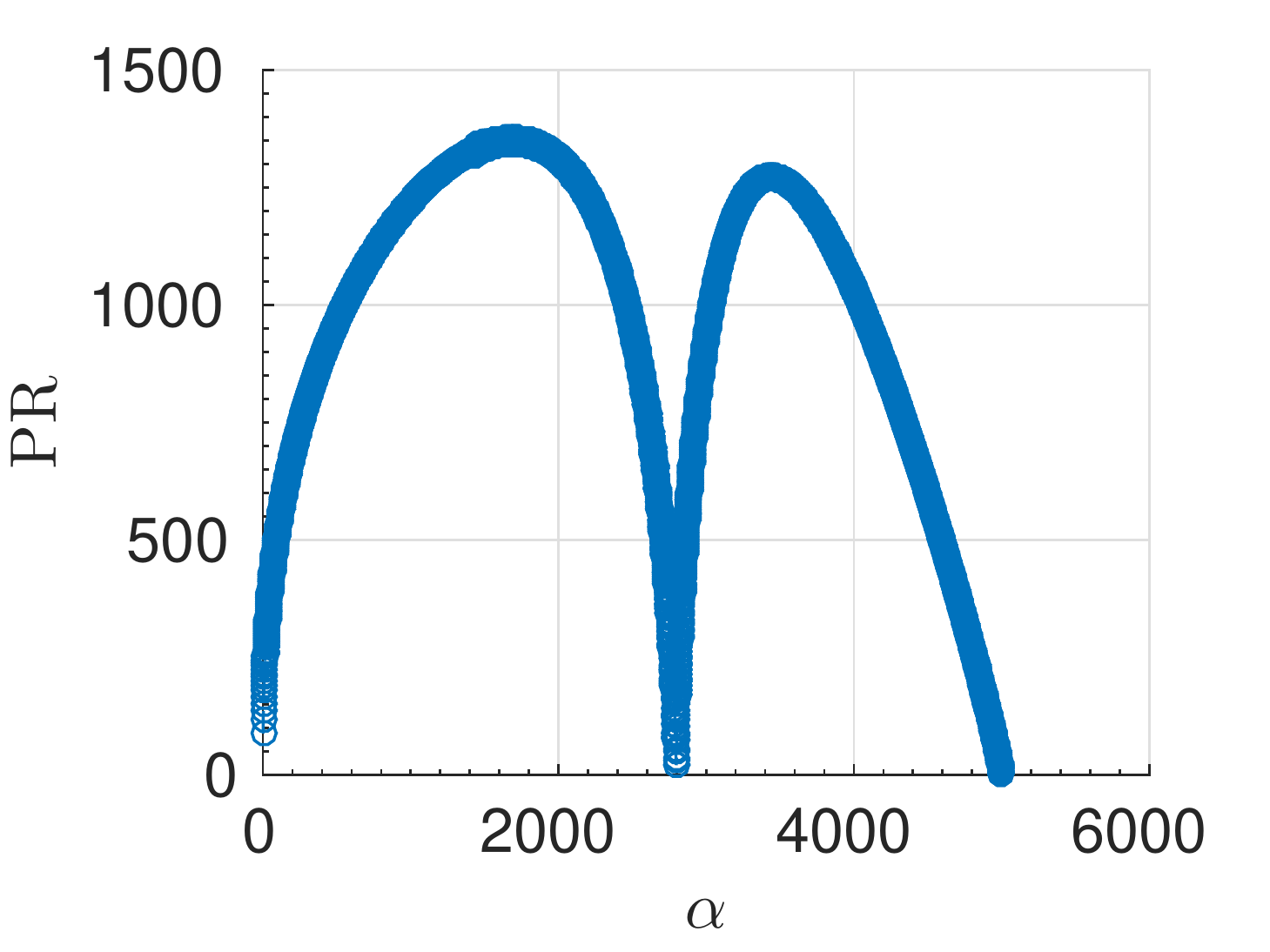}}\hfill \subfloat[]{\label{fig:fig4b}%
\includegraphics[width=0.24\textwidth]{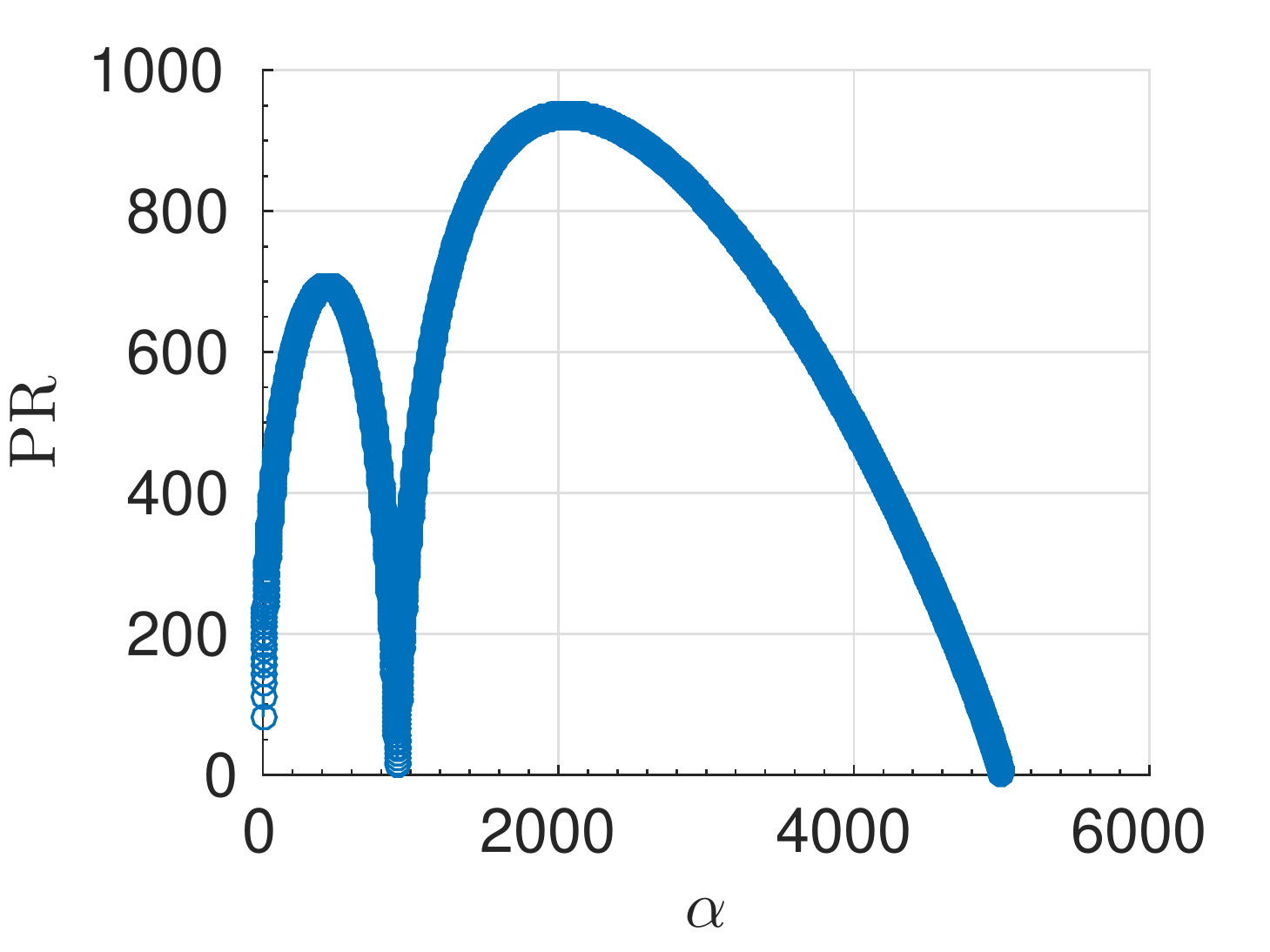}}
\caption{The participation ratio values of the eigenspectrum for a ferromagnetic
Hamiltonian with (a) $q_{f}=0.5$ and (b) $q_{f}=2$ for a particle number of $%
10^{4}$. The eigenstates are ordered ascending in energy from the ground state ($\alpha = 1$) to the most excited state.}
\label{fig:fig4}
\end{figure}

In order to understand why there is a nonlinear structure in the EEV plot,
which basically results in a nonthermal behaviour in the dynamics, we
compute other quantities which can provide more information on the
eigenspectrum structure of the model. The spinor Hamiltonian in Eq.\ %
\eqref{hamiltonian} can actually be mapped to a single quantum-particle
Hamiltonian with nearest-neighbor hopping and onsite potentials on a finite lattice. The Fock
basis $\Ket{N_{-1},N_0,N_{1}} = \{ \Ket{0,N,0},\Ket{1,N-2,1}, \cdots, %
\Ket{N/2,0,N/2} \}$ with zero total magnetization in our spinor Hamiltonian
can be mapped to a basis of different lattice sites in the language of a single
hopping particle in 1D lattice. Then the interaction terms $a_0^{\dagger} a_0^{\dagger} a_1
a_{-1} $ and $a_1^{\dagger} a_{-1}^{\dagger} a_0 a_{0}$ realize the
nearest-neighbor hopping as can be seen when we do the operation $a_1^{\dagger} a_{-1}^{\dagger} a_0 a_{0} \Ket{0,N,0} = \sqrt{N (N-1)} \Ket{1,N-2,1}$. The rest of the terms in Eq.\ %
\eqref{hamiltonian} impose an onsite potential. The tight-binding Hamiltonian for the mapping could be stated as
\begin{eqnarray}
H_{m} &=& \sum_{i=1}^{N/2-1} J(i) \left( c_{i+1}^{\dag} c_i + \text{h.c.} \right) + \sum_{i=1}^{N/2} \eta(i) c_i^{\dag} c_i, \label{mappedHamiltonian}
\end{eqnarray}
where $J(i)$ are real hopping coefficients that are a function of site position and $\eta(i)$ are the onsite potentials that depend on the site positions as well. The lattice size $N/2$ is the dimension of the Fock space. Here the exact dependence of $J$ and $\eta$ parameters on the positions of the sites in our imagined lattice is determined through the terms in the spinor BEC Hamiltonian Eq. \eqref{hamiltonian}. See App. B for how a spinor Hamiltonian engineers the lattice parameters for the mapped Hamiltonian Eq. \eqref{mappedHamiltonian}. This mapping reminds us of the physics of Anderson localization \cite{PhysRev.109.1492}, albeit the onsite potentials $\eta(i)$ are not random. Hence, we study the participation ratio (PR) 
\begin{eqnarray}
P_{\alpha} &=& \left(\sum_{n=1} |\psi_{\alpha n}|^4\right)^{-1},
\end{eqnarray}
to analyze the localization properties of the eigenstates \cite%
{RevModPhys.80.1355,haake2001quantum, PhysRevB.83.094431}; here, $\alpha$
denotes each eigenstate and $n$ is the Fock basis vectors. As seen in Fig.\ %
\ref{fig:fig4}, PR has a dip around the eigenstate corresponding to the nonthermal kink
eigenstate in its corresponding EEV plot, which points to lower PR values of the nonthermal states in the Fock basis when compared to other eigenstates in the spectrum. This result hints at a link between the nonthermal behaviour that we observe in the system and the Anderson-like localization 
\cite{PhysRev.109.1492} of the eigenstates in the Fock space. In other words, the nonthermal states of the system also seem to be the most localized states in the spectrum (excluding the edges).
\begin{figure}[tbp]
\centering
\subfloat[]{\label{fig:appDFig1a}\includegraphics[width=0.5%
\textwidth]{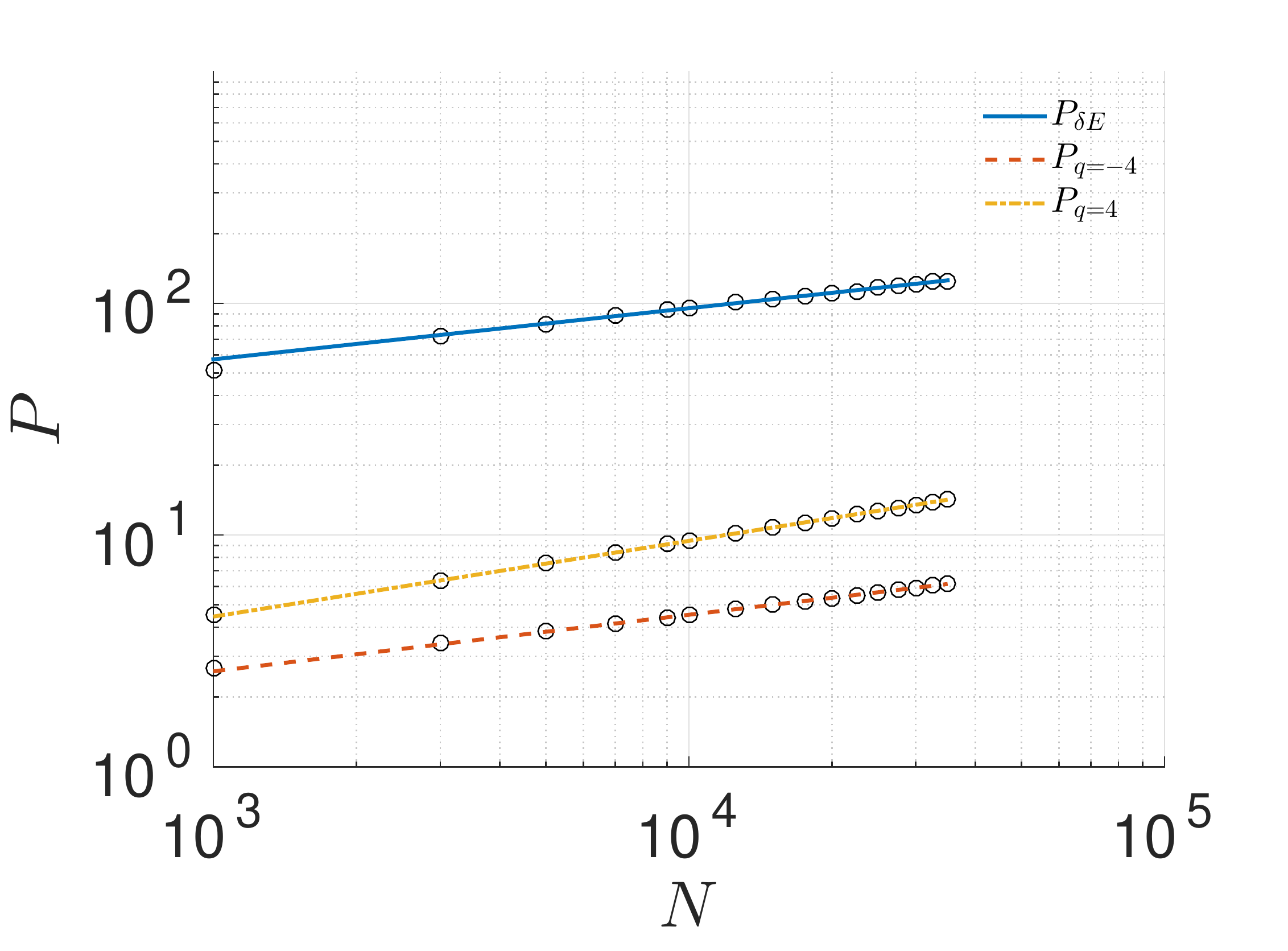}}\hfill \subfloat[]{%
\label{fig:appDFig1b}\includegraphics[width=0.5%
\textwidth]{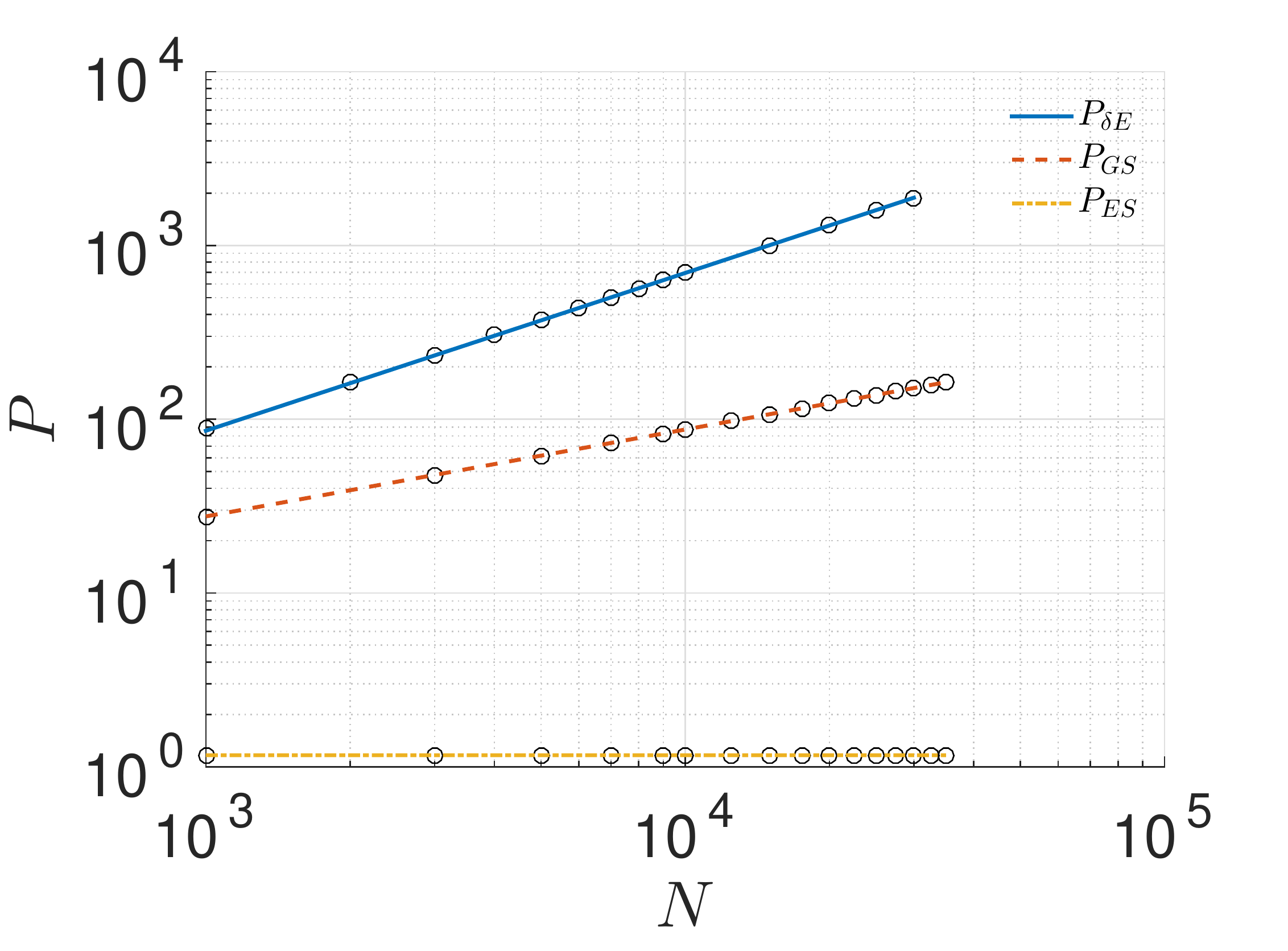}}
\caption{(Color online) System size scaling of (a) averaged participation ratio of low-PR eigenstates with a fixed energy interval of $\sim 25 [c_1]$ around the most outlier (kink) eigenstate (solid blue), the ground state participation ratio at $q=-4$ (dashed red) and at $q=4$ (dotted-dashed orange); (b) averaged participation ratio of high-PR eigenstates with a fixed energy interval of $\sim 60 [c_1]$ around the center of the spectrum (solid blue) when $q=3$ is chosen, PR of ground state (dashed red) and of the most-excited state (dotted-dashed orange) when the system is not going through one of its phase transition points e.g. $q=1$.}
\label{fig:appDFig1}
\end{figure}

In order to make this point stronger, we analyze the system size PR scaling
of eigenstates with high- and low-PR values. To target the low-PR region of the spectrum, we utilize two different methods. We emphasize that low-PR region of the spectrum in Figs. \ref{fig:fig4} (excluding the edges of the spectrum) is also the nonthermal region as already shown with the ETH indicators. There is a rapid change around the kink state which is always the extremum point of the EEV (Figs. \ref{fig:fig3c}-\ref{fig:fig3d}) and the level spacings (Fig. \ref{fig:fig2c}). Additionally the kink state slightly shifts in the spectrum as we increase the system size upto thermodynamic limit. So, even though we are able to detect the kink state in the spectrum with all these observations, we note that the kink state shows consistently low PR values for each system size but its scaling is not well-defined possibly due to finite-size effects. Therefore, the first method we apply is averaging over low-PR states around the most outlier (kink) state for each system size with a fixed energy interval. The solid line in Fig. \ref{fig:appDFig1a} is the scaling behaviour that we observe for this method when $q=-0.65$ is chosen, which is also a $q$ value that keeps the kink state around the center of the spectrum. The extracted scaling exponent is $\gamma \propto 0.22$ with $R^2 = 0.997$. The second method employs the phase transition points. We know that the ground state is the kink eigenstate at phase transition points when $q=4$ or $q=-4$ is taken in the thermodynamic limit. Even though for a finite size condensate the phase transition points are slightly off from $q=4$ and $q=-4$ and hence the ground state is not exactly the kink eigenstate, the region around the ground state is the nonthermal kink region. This observation can be made through the difference in the PR scaling exponents of the ground state when we have $q=4$ (or $q=-4$) and $q$ is away from the phase transition points. Fig. \ref{fig:appDFig1b} dashed line shows the scaling of the ground state when $q=1$ which we extract $P \propto N^{0.5}$ ($R^2 = 1$). The exponent $\gamma \propto 0.5$ is obtained for any $q$ sufficiently far from  $q=4$ or $q=-4$. On the other hand, we obtain a scaling of $P \propto N^{0.32} (R^2 = 1)$ and $P \propto N^{0.24} (R^2 = 0.999)$ for $q=4$ and $q=-4$, respectively. Thus, clearly the ground state is neither localized nor extended completely when the system is not at its phase transition points. However, when the system goes through its phase transitions, the ground state coincides with the low-PR region states and this provides us a way to estimate the scaling exponent of states at the low-PR region. We note that extracting a well-defined scaling only for the most outlier (kink) state in a finite-size system is difficult, but still averaging over a couple of states around it gives an idea about the localization properties of the nonthermal region. Overall the extracted scaling exponents point out to that the low-PR nonthermal kink region is not completely localized region with a scaling exponent of $\gamma = 0$, however it is the most localized region of the spectrum. The high-PR eigenstates that are also responsible of thermalization observed in the system show a scaling of $P_{\delta} \propto N^{0.91}$ ($R^2 = 1$), when we choose a fixed energy interval in the middle of the spectrum for Zeeman field strength $q=3$, (solid line in Fig. \ref{fig:appDFig1b}). We observe almost the same PR scaling with exponent $\gamma = 0.9$ for single eigenstates chosen at the high-PR section of the spectrum and for different $q$ values. Even though such an eigenstate is not completely extended with a scaling exponent of $\gamma = 1$, it is the most extended region of the spectrum. All in all, the previous analysis of the ETH indicators clearly distinguishes the thermal and nonthermal states in the system and PR analysis demonstrates a link between localization and thermalization properties of our system, even though the thermal and nonthermal states are not completely delocalized and localized, respectively.

Finally, we note the difference between the behaviours seen in regions I and IV. Although the equilibrium behaviour in region IV can be predicted by microcanonical ensemble as seen in Fig. \ref{fig:thermalization2}, its cause is not related to the eigenstate localization properties. We observe almost constant dynamic evolution (or almost-no nonequilibrium evolution) for the simulations at this section, which implies that one of the eigenstates dominates the evolution. In the case seen in Fig. \ref{fig:thermalization2}, it is the most-excited state that
governs the dynamics for negative $q_f$ values. The most-excited state shows
a constant PR scaling with an exponent of $0$ (dashed-dotted line in Fig. \ref{fig:appDFig1b}). So, even though the
eigenstate is perfectly localized, the initial state is already in equilibrium with the quench Hamiltonian, which leads to the thermalization. In Fig. \ref{fig:thermalization2}, also note that we observe thermalization
for values at $q_f > 4$ because now the initial state mostly resembles to
the ground state of the quench Hamiltonian instead of the most-excited
state. Finally, even though we show the PDE values at region III in Fig. %
\ref{fig:thermalization2}, we should remind the reader that the dynamics of
region III does not equilibrate but shows large fluctuations around its PDE value (which will be discussed in the next section as a special case).

An important difference between the spinor BEC model and the single quantum-particle hopping model is that even though the observable $\Braket{N_{0}}$ is local in
the spinor BEC case, it is a non-local observable when it is mapped onto the
particle lattice. However more importantly, our model does not translate
to an Anderson model with random potentials. Single quantum-particle hopping model
with random potentials leads to sites with very low PR values. It is also
analytically known that such a model cannot cause thermalization and satisfy
ETH \cite{PhysRevLett.116.030401}. Therefore, based on our results with
spinor BECs, we argue that engineering the potential of a single quantum-particle model should prevent the localization in the particle lattice
and give rise to thermalization for global observables defined for this
model.

\section{Existence and Absence of Quantum Collapse and Revivals}

In this section, we analyze the cases that demonstrate quantum collapse and
revivals and derive an analytical expression to predict their time scales.
Further we examine the scaling of collapse and revival times with the number
of particles in the condensate to be able to present realistic predictions
for the experiment. Finally we discuss `the special cases', where we do not
observe a revival or even equilibration. 
\begin{figure}[tbp]
\centerline{\label{fig:fig1a}\includegraphics[width=0.49%
\textwidth]{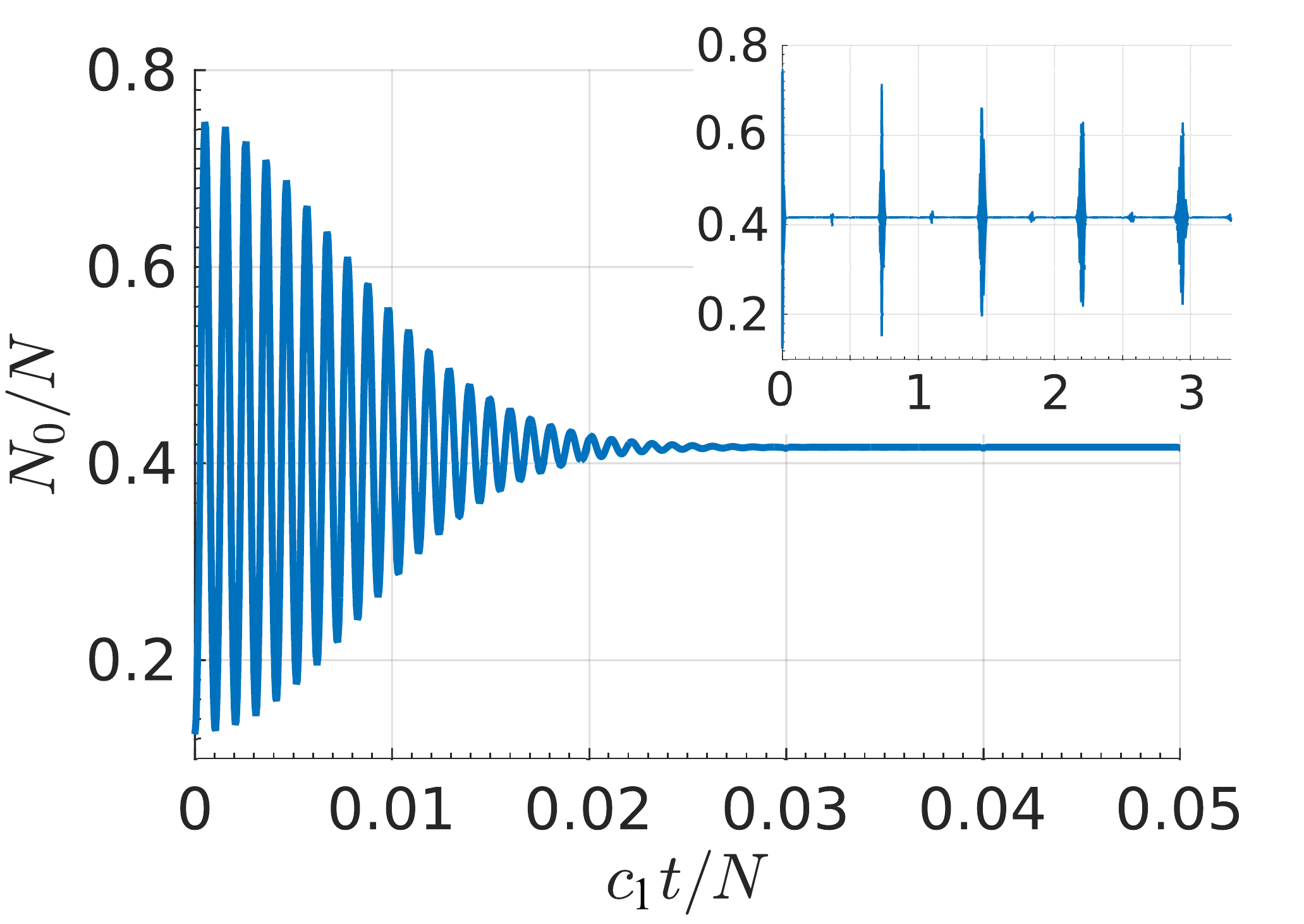}}\hfill  
\caption{(Color online) The sudden quench dynamics in the short time-scale
showing the collapse in detail when there is $N= 2\times 10^3$ particles in
the condensate and x-axis is scaled with the number of particles when we quench from $q_i=-3$ to $q_f = -0.5$ for the ferromagnetic case. The inset
plot shows the revivals in long time-scale.}
\label{fig:fig1}
\end{figure}
\begin{figure}[tbp]
\centering
\subfloat[]{\label{fig:fig2a}\includegraphics[width=0.24%
\textwidth]{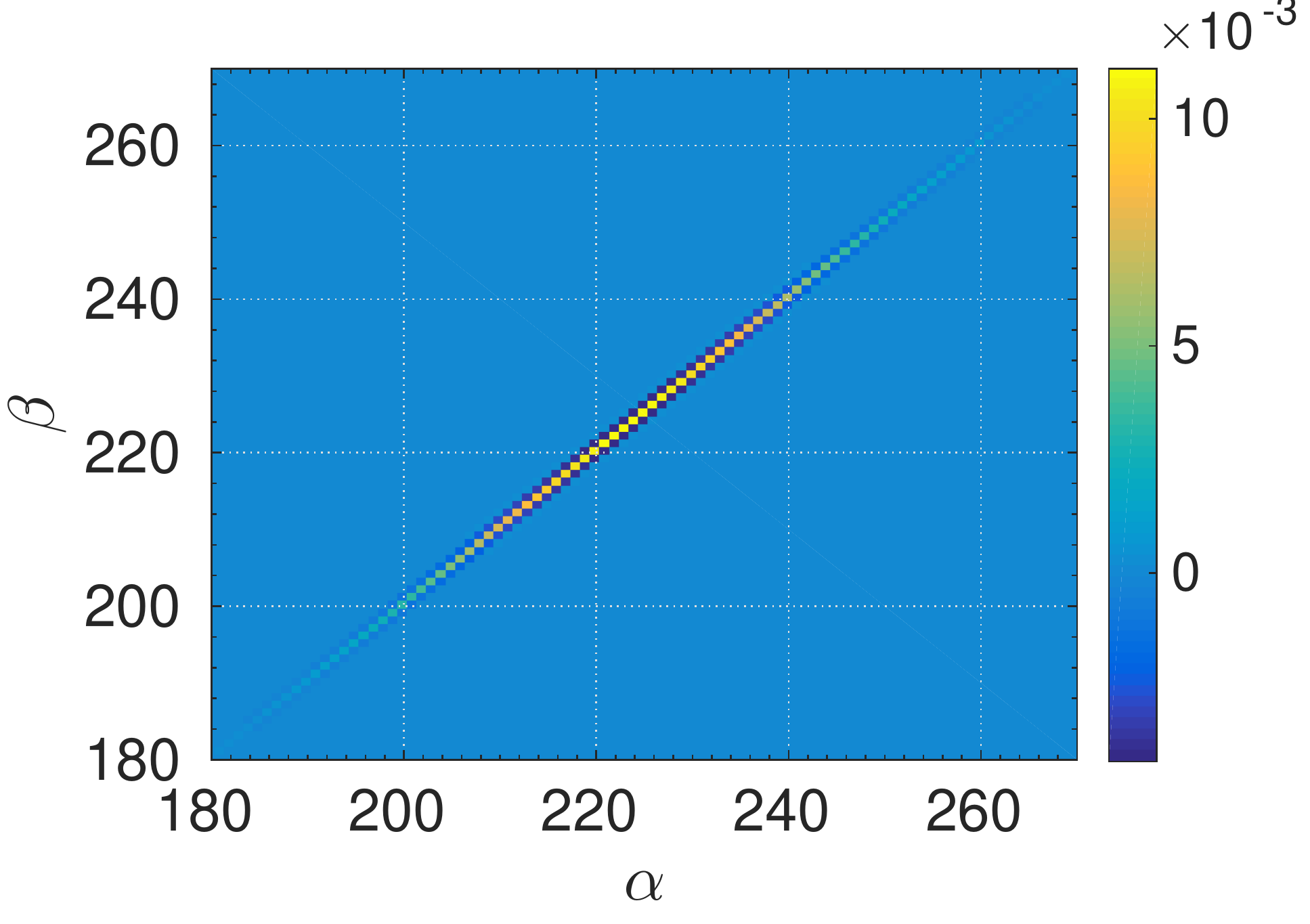}}\hfill \subfloat[]{\label{fig:fig2b}%
\includegraphics[width=0.24\textwidth]{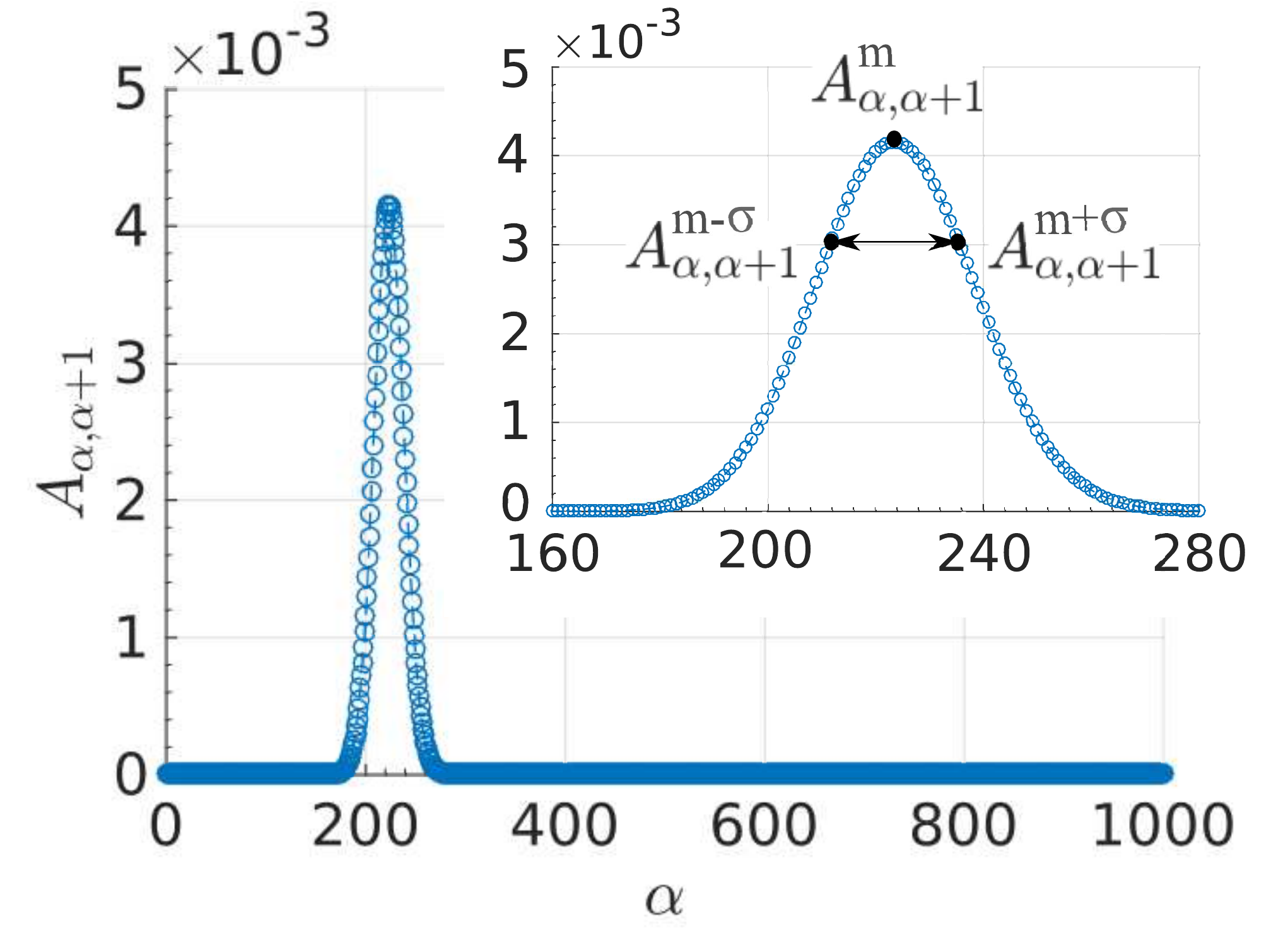}}
\par
\medskip \subfloat[]{\label{fig:fig2c}\includegraphics[width=0.23%
\textwidth]{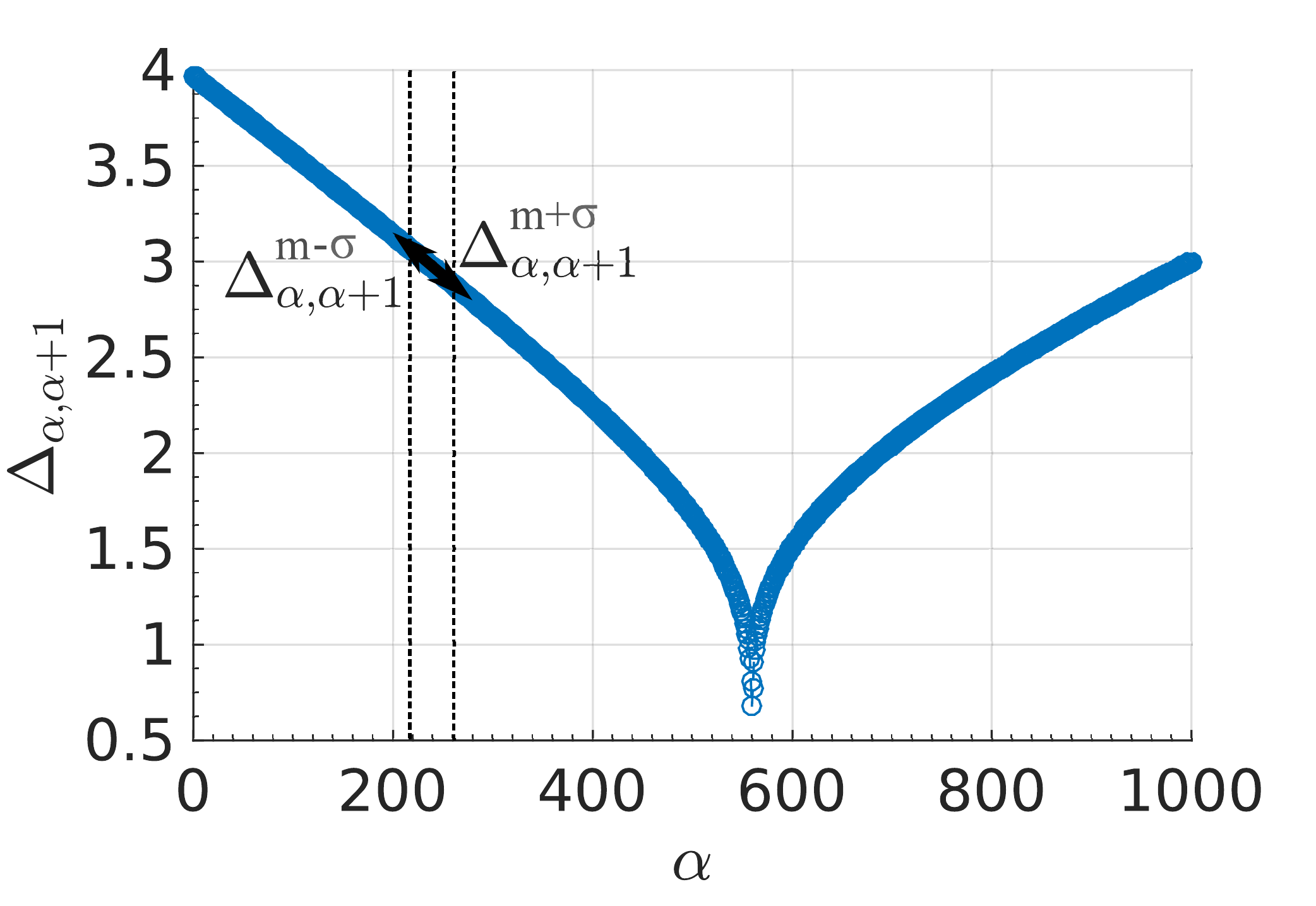}}\hfill 
\subfloat[]{\label{fig:fig2d}
\includegraphics[width=0.23\textwidth]{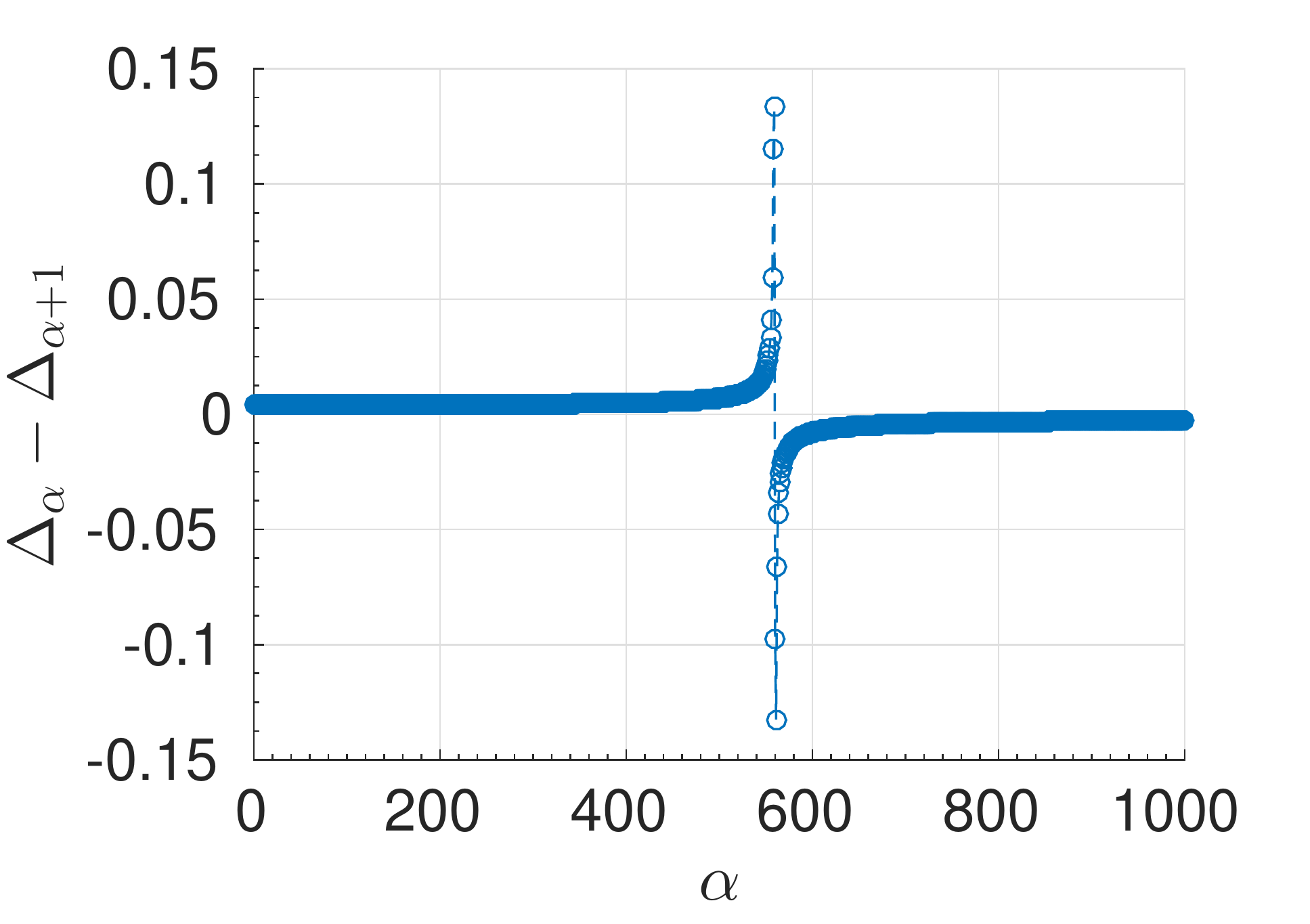}}
\caption{(Color online) (a) The overlap matrix with respect to eigenstates $\alpha$ and $\beta$, (b) the first off-diagonal
terms of the overlap matrix, (c) the nearest-neighbor (NN) energy gaps and
(d) The difference/derivative plot of the NN energy gaps with $N= 2\times
10^3$ particles in the ferromagnetic condensate for the quench from $q_i = -3$ to $q_f = -0.5$. The x-axis is the eigenstates $\alpha$ ordered ascending in energy from the ground state $\alpha=1$ to the most-excited state.}
\label{fig:fig2}
\end{figure}
%%ST: 1. label the t_{c} and t_{r} or what they correspond in figure (b), or make a smaller inset of (b) for that? Basically, I think a label will be much easier to see than your formula. 2. (a) has \alpha and \beta on the axis instead? 3. in (d), your notation on the y-axis is not defined (different from (c)). 

Now we choose a point on the ferromagnetic sudden quench map Fig.\ \ref{fig:PD1} that
thermalizes which can be detected via Figs.\ \ref{fig:thermalization1} and %
\ref{fig:thermalization2}. So then, if we quench from $q_i = -3$ to $q_f =
-0.5$, we observe a series of collapse and revivals in Fig.\ \ref{fig:fig1},
and the equilibration value in between matches both diagonal and
microcanonical ensembles. A collapse before equilibration is what mostly
observed in experiments. We also intuitively expect to see a series of
revivals due to the finite-size effects. However in order to understand how
collapse and revivals emerge in our model, let us go back to the sudden
quench procedure given in previous section and modify the Eq. (\ref{eq1}). 
%%ST: what do you mean here? Why did sudden quench predicts thermalization? The earlier plots show diagonal ensemble matches microcanonical ensemble, but it does not imply the dynamics can reach the equilibrium (diagonal ensemble). Maybe rephrase it in some way.  
Notice that $c_{\alpha}^* c_{\beta} = c_{\beta}^* c_{\alpha}$ when the
coefficients are real, which is the case in our problem. Also $N_{0,\alpha
\beta} = N_{0,\beta \alpha}$ in our model. Then we can regroup Eq.\ (\ref%
{eq1}) as, 
\begin{eqnarray}
\Braket{N_0(t)} &=& \sum_{\alpha \geq \beta} c_{\alpha}^* c_{\beta}
\left(e^{i (E_{\alpha}-E_{\beta}) t}+e^{-i (E_{\alpha}-E_{\beta}) t}\right)
N_{0,\alpha \beta},  \notag \\
&=& 2 \sum_{\alpha \geq \beta} c_{\alpha}^* c_{\beta} \cos
\left((E_{\alpha}-E_{\beta})t\right) N_{0,\alpha \beta}.  \label{eq2}
\end{eqnarray}
Eq.\ (\ref{eq2}) tells us that dynamics we observe in a sudden quench is the
interference of sinusoidal functions weighted with some overlap. We can
write Eq.\ (\ref{eq2}) more clearly as, 
\begin{eqnarray}
\Braket{N_0(t)} &=& \sum_{\alpha \geq \beta} A_{\alpha \beta} \cos
\left(\Delta_{\alpha \beta} t\right),  \label{eq3}
\end{eqnarray}
where $\Delta_{\alpha \beta} = E_{\alpha}-E_{\beta}$ is the energy gaps, 
%%ST: why nearest neighbor? Everything comes in here. The nearest-neighbor dominates because of A_{\alpha \beta}.
Fig.\ \ref{fig:fig2c} and $A_{\alpha \beta} = 2 c_{\alpha}^* c_{\beta}
N_{0,\alpha \beta}$ is the overlap matrix, Fig.\ \ref{fig:fig2a}. We note
that the diagonal terms are the most populated terms in the overlap matrix
and they correspond to the diagonal ensemble prediction. In fact it is
important that the off-diagonal terms vanish for thermalization to happen or
they should be much smaller compared to diagonal terms. We observe this is
almost the case in Fig.\ \ref{fig:fig2a}, except the first and second
off-diagonals still contribute to the dynamics even though they are much
smaller than the diagonal terms. Fig.\ \ref{fig:fig2b} shows the first
off-diagonals of the overlap matrix (which we call overlap distribution in the following). This Poisson-like overlap exists when the dynamics demonstrate a series of collapse and revivals and it turns out
to be important in determining the time scales of collapse and revivals in
spinor condensates under SMA.

The time scale of a collapse is related to the time when the oscillating
terms with an energy gap argument in Eq.\ (\ref{eq3}) start to become
uncorrelated. The terms corresponding to the farthest ends of the
distribution are also the farthest in oscillation frequency. They become
uncorrelated after all the other terms get uncorrelated. From that point on,
all the oscillating terms will be destructively interfering. We estimate
these elements with root-mean-square of the overlap distribution as also
done for collapses in Jaynes-Cummings model \cite{scully1997quantum}. Ref. 
\cite{PhysRevLett.54.1879} predicts the collapse time for the Ising model as
inversely proportional to the energy spread of the initial state, which is
similar to our criteria and expression. The following collapse time expression produces a value of $c_1 t_c/N \sim 0.02$ for the quench simulation depicted in Fig. \ref{fig:fig1}:
\begin{eqnarray}
t_c &=& \frac{2\pi}{|\Delta_{\alpha, \alpha+1}^{m+\sigma}-\Delta_{\alpha,
\alpha+1}^{m-\sigma}|},
\end{eqnarray}
where $\Delta_{\alpha, \alpha+1}^{m}$ denotes the nearest-neighbour energy gap (level spacing, Fig. \ref{fig:fig2c}) corresponding to the maximum value in the overlap distribution (Fig. \ref{fig:fig2b}) and hence $\Delta_{\alpha, \alpha+1}^{m+\sigma}$ is the nearest-neighbour energy gap corresponding to the value which is $\sigma$
farther from the mean in the distribution (cf. the inset of Fig. \ref{fig:fig2b}). It is possible to fine-tune the predicted collapse
time by taking more than $1\sigma$ of the overlap distribution. Also note that we find $c_1 t_c \sim N^{1/2}$ as the scaling of the collapse time-scale.

A quantum revival happens when all the oscillating terms become correlated
with each other again. This can be measured through the difference between
nearest-neighbour energy gaps corresponding to the the mean $\Delta_{\alpha, \alpha+1}^m$ and the closest
point to mean $\Delta_{\alpha,\alpha+1}^{m-1}$ in the overlap distribution (cf. the inset of Fig. \ref{fig:fig2b}), 
\begin{eqnarray}
t_r &=& \frac{2\pi}{|\Delta_{\alpha, \alpha+1}^m-\Delta_{\alpha,\alpha+1}^{m-1}|}.
\label{revivalTS}
\end{eqnarray}
Fig.\ \ref{fig:fig2d} shows the differences between nearest-neighbour energy
gaps. Note that $\Delta_{\alpha} - \Delta_{\alpha+1}$ are mostly flat around
where the overlap distribution is nonzero. This is vital for a collective revival to occur, since otherwise terms in Eq. \eqref{eq3} will never constructively interfere at a fixed time, namely the revival time. When we
have $t_{r} (\Delta_{\alpha} - \Delta_{\alpha+1}) = 2\pi$, all oscillating terms interfere constructively, creating the first revival. Both the analytical expression and the data analysis give a revival time $%
c_1 t_r/N \sim 0.735$. Since the scaling of the revival time-scale turns out to be $c_1 t_r \sim N$, this value can be obtained for all sizes for the parameters depicted in Fig. \ref{fig:fig1}. Also note that the linearly growing recurrence times is well-known in the literature \cite{eisert2015quantum}. The small peaks between the collapse and revivals seen in Fig.\ \ref{fig:fig1a} are the small revivals contributed by the second off-diagonal terms in the overlap matrix, Fig.\ \ref{fig:fig2a}. We can also predict the oscillation frequency, 
\begin{eqnarray}
t_{\text{osc}} &=& \frac{2\pi}{\Delta_{\alpha,\alpha+1}^m},
\end{eqnarray}
by using the nearest-neighbor energy gap at the maximum point of the overlap
distribution $\Delta_{\alpha,\alpha+1}^m$. There is an another interesting quantity that can be predicted
in a collapse-revival picture. We observe how revivals are suppressed in a
very long time scale in the inset of Fig.\ \ref{fig:fig1a}. This `randomizing time' is
where the initial memory of the system irreversibly gets lost. Even though a
typical randomizing time is out of experimental reach, it is interesting to
note that an isolated, unitary and finite-size quantum system will be
eventually randomized and hence completely thermalized at the randomizing
time which can be estimated via 
\begin{eqnarray}
t_{rz} &=& \frac{2\pi}{|\Delta^{\prime }_{m+\sigma}-\Delta^{\prime}_{m-\sigma}|},
\end{eqnarray}
where $\Delta^{\prime }= \Delta_{\alpha}-\Delta_{\alpha+1}$ denotes the
difference between nearest-neighbor energy gaps (Fig. \ref{fig:fig2d}) and the rest of the notation is same with the previous definitions where we use the overlap distribution for $m \pm \sigma$. 

In order to give a sense of these time scales, let us fix the particle
density in our condensate to $5\times 10^{14}$ cm$^{-3}$. Then the
coefficient reads $c_1 \sim -2\pi \times 9$ Hz, which gives a realistic
collapse time of $\sim 0.5$ s and a revival time of $\sim 25$ s for a
condensate particle number of $2\times 10^3$. 
\begin{figure}[tbp]
\centering
\subfloat[]{\label{fig:fig9a}\includegraphics[width=0.49%
\textwidth]{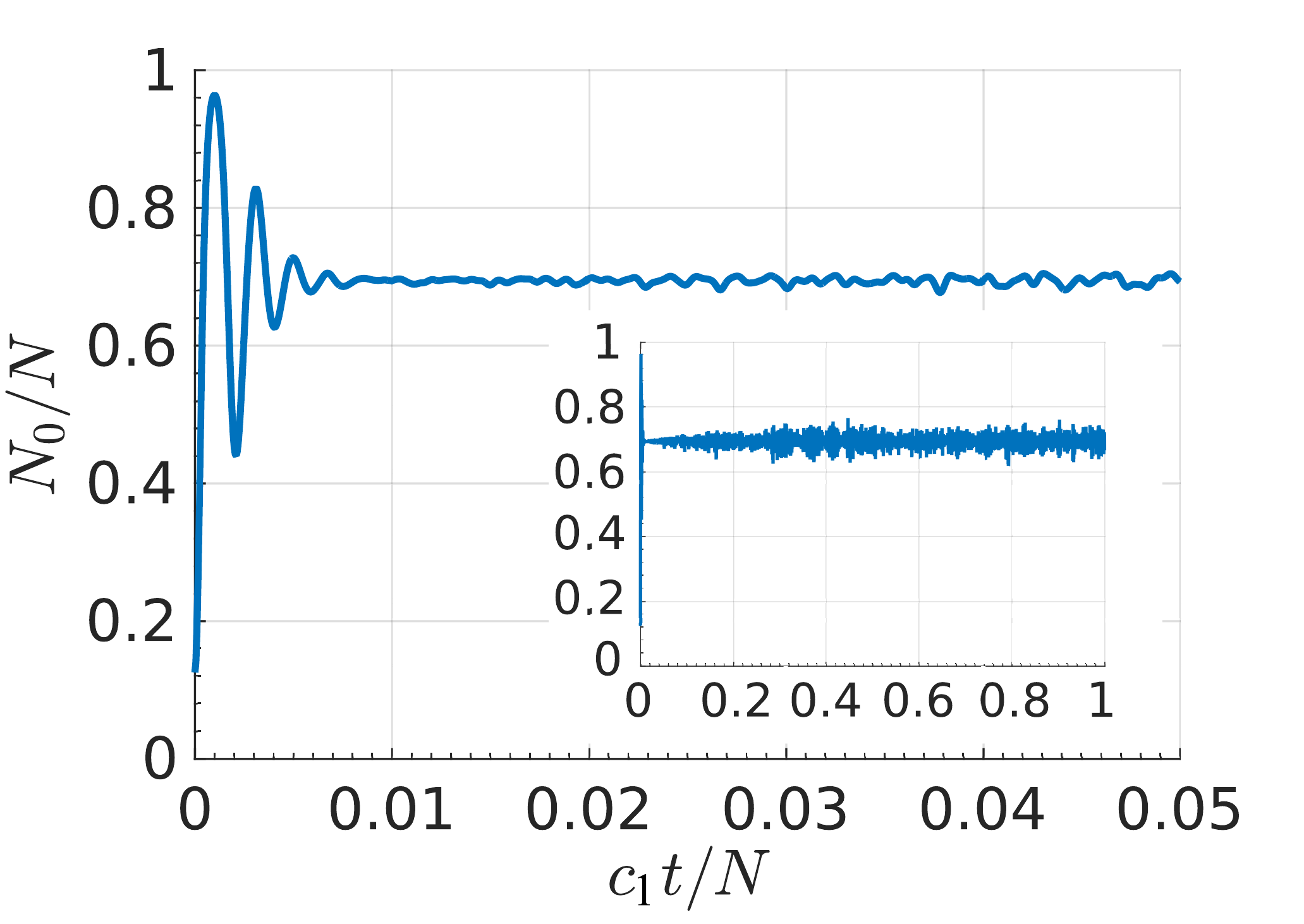}}\hfill \subfloat[]{\label{fig:fig9b}%
\includegraphics[width=0.49\textwidth]{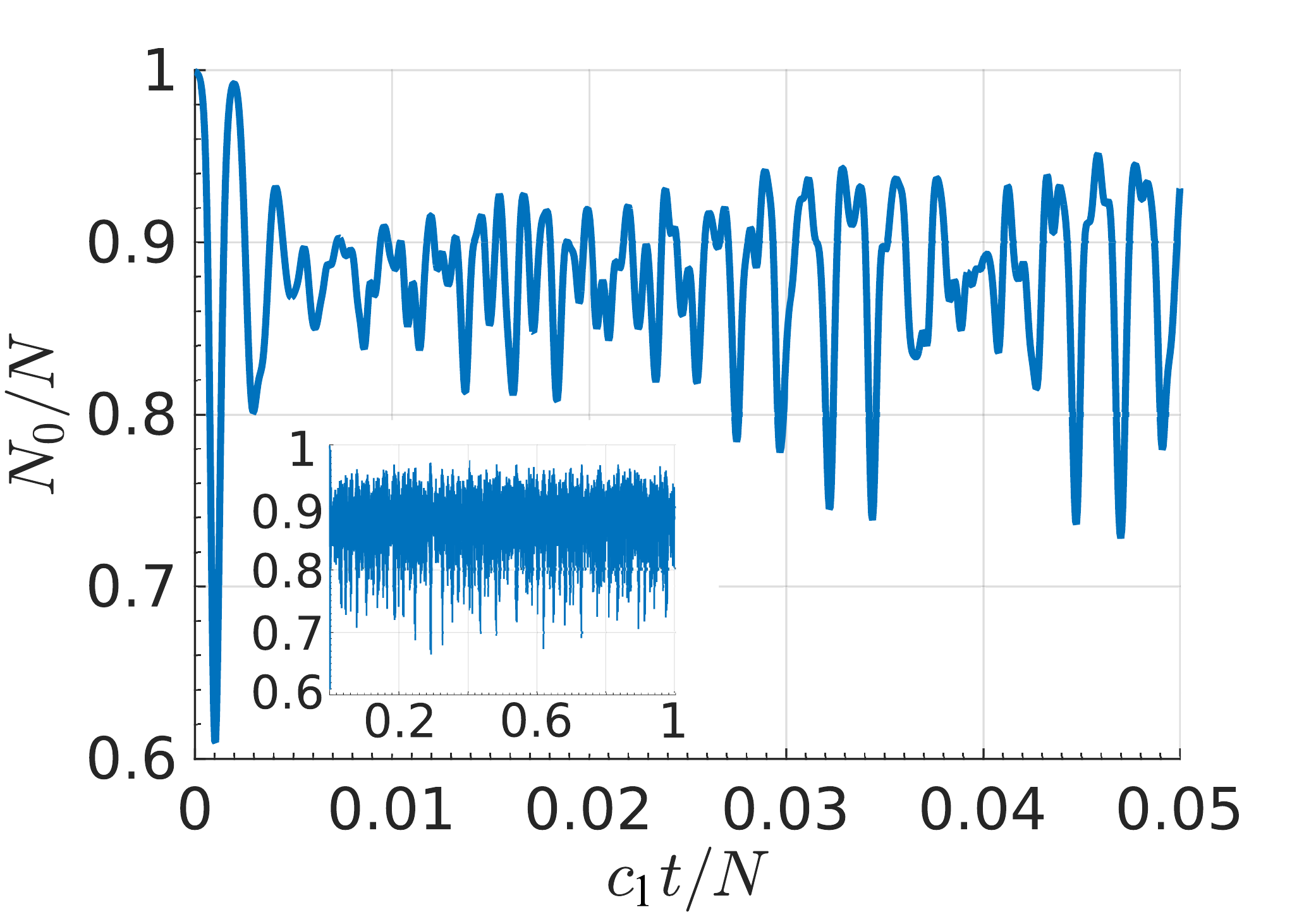}} 
%\subfloat[]{\label{fig:fig2c} \includegraphics[width=0.5\textwidth]{figures/dif_delta.eps}} 
\caption{(Color online) The sudden quench dynamics in short time-scale (a)
from $q_i = -3$ to $q_f = 0.5$ and (b) from $q_i = 4.1$ to $q_f = 2$ with
insets of long time-scales for ferromagnetic condensates and $N= 2\times 10^3$ particles.}
\label{fig:fig9}
\end{figure}
This sudden quench experiment corresponds to a data point on Fig.\ \ref%
{fig:thermalization1}, where ETH can explain the match between the thermal
relaxation values predicted by diagonal ensemble and microcanonical
ensemble. Therefore we can conclude that there is thermalization until the
initial memory of the system comes back with a quantum revival. 
\begin{figure}[tbp]
\centering
\subfloat[]{\label{fig:appEFig1a}\includegraphics[width=0.24%
\textwidth]{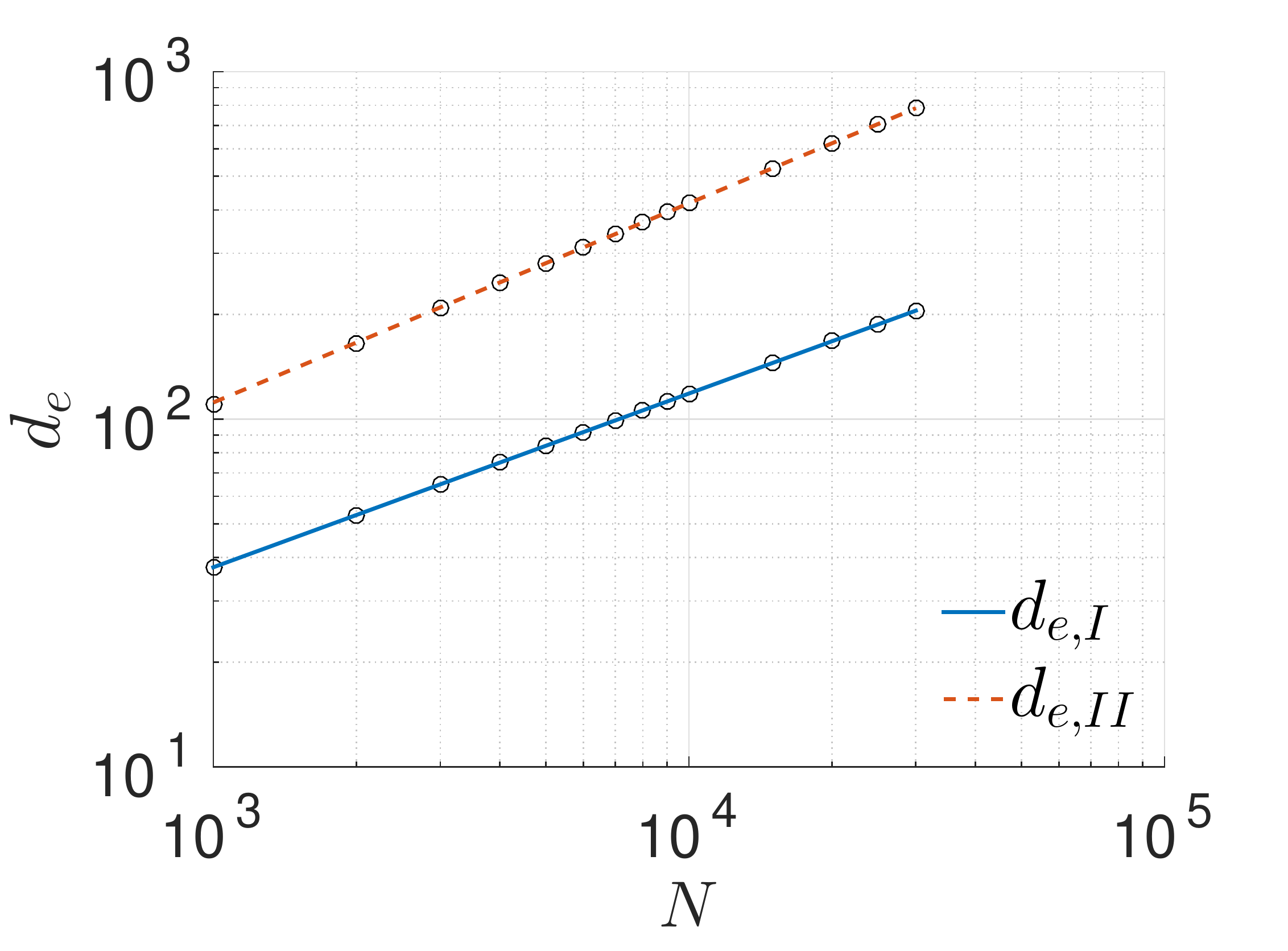}}\hfill \subfloat[]{%
\label{fig:appEFig1b}\includegraphics[width=0.24%
\textwidth]{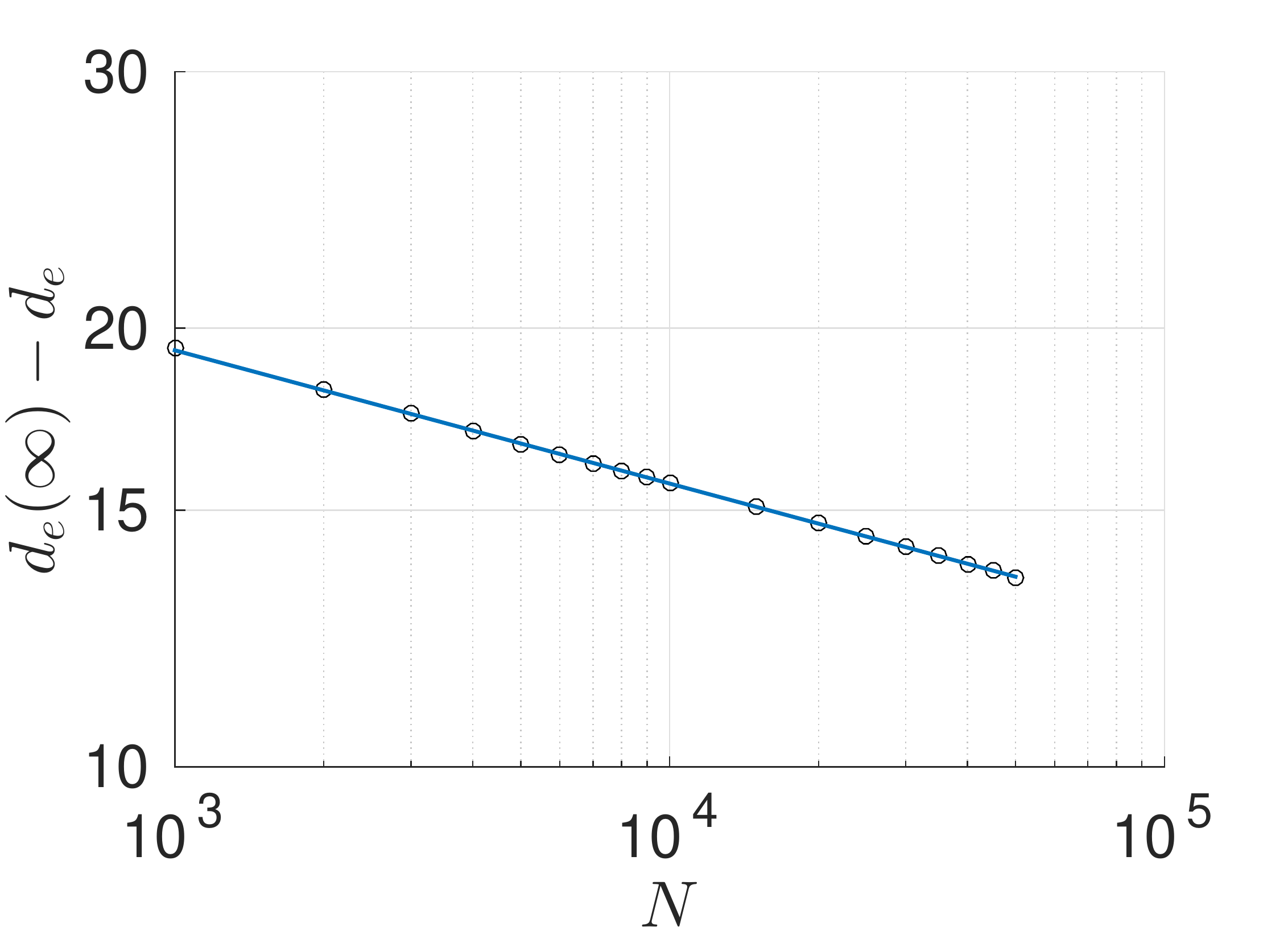}}
\caption{(Color online) The effective dimension scaling for a quench (a) from $q_i = -3$ to $q_f = 0.5$ (Region II) (dashed red) with $d_e \propto N^{0.57}$, from $q_i = -3$ to $q_f = -0.5$ (Region I) (solid blue) with $d_e \propto N^{0.5}$ and (b) from $q_i = 4.1$ to $q_f = 2$ (Region III)  with $d_e = 28.3 - 36.3 N^{-0.092}$ with respect to system size. The correlation coefficient is $R^2 = 1$ for all figures. $d_e(\infty)$ stands for the offset value of the fitting in subfigure b.}
\label{fig:appEFig1}
\end{figure} 
Then it is important to see how the times of the collapse and the first
revival scale with the number of particles in the condensate. Reminding the reader of $c_1
t_c \sim N^{1/2}$ and $c_1 t_r \sim N$ and using the estimations done for
Thomas-Fermi limit in Ref.\ \cite{PhysRevA.60.1463}, we figure out that $c_1
\sim N^{2/3}$ in 1-dimension, hence $t_c \sim N^{-1/6}$ and $t_r \sim
N^{1/3} $. Although SMA breaks down in large condensate limit \cite%
{PhysRevA.60.1463} and the experiments always have finite sizes, it is still
insightful to imagine the thermodynamic limit $N \rightarrow \infty$. In
thermodynamic limit, a 1D spinor BEC system has a diverging revival time and
a vanishing collapse time, which implies thermalization described by ETH for
our model.

Now let us choose a point on the map Fig.\ \ref{fig:PD1} that does not
thermalize to illustrate one of the special cases. If we quench from $q_i =
-3$ to $q_f = 0.5$ (corresponding to the parameters in Fig.\ \ref{fig:fig3}a
and \ref{fig:fig3}c), we observe the dynamical behaviour in Fig.\ \ref%
{fig:fig9a}. There is a well-defined collapse whose time-scale can be
predicted with the collapse criterion and the system seems to equilibrate right after the collapse. However looking at the dynamics for a longer time (inset of Fig.\ \ref{fig:fig9a}) reveals that the revivals attempt to happen at different times resulting with no collective recurrence for a finite system. This is due to the broad shape of the EON window (Fig.\ \ref{fig:fig3}a). One can calculate the so-called effective dimension of the system \cite{PhysRevLett.101.190403,1367-2630-13-5-053009} under this specific quench, which is the participation ratio of the initial state in the eigenstate reference basis instead of Fock basis,
\begin{eqnarray}
d_e &=& \left( \sum_{\alpha} |c_{\alpha}|^4 \right)^{-1},
\end{eqnarray} 
where $|c_{\alpha}|^2$ is the eigenstate occupation numbers as in Eq. \eqref{eqInitState}. The effective dimension is a measure of how broad the EON window is. 
%We see that the effective dimension of the quench from $q_i =-3$ to $q_f = 0.5$ is a big enough value $d_e \gg 1$ for the system to equilibrate eventually. However 
In order to determine if a quantum system equilibrates, one needs to look at the scaling of the effective dimension with the system size. We find a scaling of $d_e \propto N^{0.57}$ ($R^2 = 1$) for this quench (Fig. \ref{fig:appEFig1a}) and in fact almost the same exponent for any other quench in region II of the sudden quench map. Therefore, we argue that in thermodynamic limit the effective dimension diverges $d_e \rightarrow \infty$ as $N \rightarrow \infty$, which leads the system to equilibration. For a comparison with region I, we calculated the effective dimension of a region I quench from $q_i =-3$ to $q_f = -0.5$ which is already shown to thermalize and hence equilibrate. As seen in Fig. \ref{fig:appEFig1b}, the effective dimension is found to be $d_e \propto N^{0.5}$ ($R^2 = 1$) and this scaling exponent is universal for all the quenches in region I. Hence, the previous argument follows. If we return to the discussion on region II dynamical behaviour, the overlap distribution (first off-diagonals in the overlap matrix) is similar in shape with Fig.\ \ref%
{fig:fig3a}. Further computations show that the energy gap differences
between neighbouring terms in the overlap distribution are different and
hence they give rise to different revival times (see. Eq. \ref{revivalTS} and Fig. \ref{fig:fig2d} around the kink region) confirming the dynamical
response. Also clearly the EON for this point on the map (Fig.\ \ref%
{fig:fig3a}) is not narrow enough to avoid the kink nonthermal states, which
causes nonthermalization for the system. As a result, the system only
equilibrates with no collective recurrence for any finite dimensions of the system. Also note that as we increase the system size, the time-scale of the revival attempts diverges which leaves us with the equilibrated section seen right after the decay. This is the behaviour that we observe for the region II on sudden quench
map Fig. \ref{fig:PD1}.

The second special point on the map Fig.\ \ref{fig:PD1} is a quench from $%
q_i = 4.1$ to $q_f = 2$, which demonstrates the behaviour for region III on
Fig. \ref{fig:PD1}. Fig.\ \ref{fig:fig9b} shows oscillatory behaviour around the system's PDE value for all times without any collapse or revival. The overlap distribution looks like
Fig.\ \ref{fig:fig3b}, however differently the first off-diagonal terms are
not really smaller than the diagonal terms (EON of the system) and in fact second and third
off-diagonal terms in the overlap matrix $A_{\alpha \beta}$ substantially contribute to the dynamics, too. This is in fact why we observe large fluctuations (Eq. \ref{eq3}). The scaling of the effective dimension for this quench turns out to be $d_e \propto 28.3 - 36.3 N^{-0.092}$ ($R^2 = 1$), which implies that in thermodynamic limit $d_e \rightarrow 28.3$ while $N \rightarrow \infty$ and the effective dimension is going to saturate at a constant value (Fig. \ref{fig:appEFig1b}). This will lead to nonequilibration since the effective dimension will be so much smaller than the dimension of the Hilbert space, $d_e \ll d_H = \infty$. We note that all quenches on region III shows a universal scaling exponent with slightly different scaling parameters. As a final remark, the EON window is narrow enough to coincide only with the nonthermal kink states implying the PDE of the system is not the thermal prediction.

\section{Conclusions and Discussion}

%And also somehow mention the special cases even in Fig 2. That's why I commented earlier that it may be better to include in figure 2 another two plots of the standard situation. And ultimately, the special cases are when the EON overlaps with the kink right? You may want to drive this point too, and talk about the kink.  

Spinor BEC model with SMA has an eigenstate expectation value spectrum for
the observable $\Braket{N_{0}}$ (the number of particles with spin-0 component in the
condensate) that shows thermalization in the context of eigenstate
thermalization hypothesis in the weak form when the quadratic Zeeman term is 
$|q|<4$ due to the `rare' nonthermal states and in the strong form,
otherwise. We adopted widely used ETH indicators to obtain our results, e.g.
support, ETH noise (fluctuations), maximum divergence from the
microcanonical prediction for an eigenstate in a fixed energy interval and
the EEV differences. We studied the effect of these nonthermal states in the
spectrum by driving the system out of equilibrium via a sudden quench from
the ground state of an initial Hamiltonian with $q_{i}$ to a final
Hamiltonian with $q_{f}$. Even though this procedure allowed us to study
certain initial conditions, we are able to generalize the results and
predict the behaviour of the system with an arbitrary initial condition. On
the other hand, such a procedure is experimentally realizable and we have
shown that it leads to a classification scheme of the system dynamics: the
sudden quench maps, Figs.\ \ref{fig:PD1} and \ref{fig:PD2}. Sudden quench
maps give us the prediction of diagonal ensemble in the long-time limit or
the long-time average of the dynamical response. For a region when the
system does not equilibrate (e.g. Region III), the value on the map is the
average of the response.

We observed that ETH is satisfied in region I with well-defined collapse and
revivals where the revival time-scale is out of reach for realistic
condensate sizes. For the region II, the dynamics equilibrate around a
nonthermal value right after a collapse (shown via the scaling of effective dimension) due to the effect of
nonthermal rare states in the spectrum. Even though dynamics at region II shows attempts for a quantum revival, not all the oscillating terms become correlated at the same time, implying the lack of a
clear quantum revival. We interpreted the thermalization seen for the
region I as weak ETH, because even though the initial state does not
overlap with the rare nonthermal states (kink region), these states still
exist in the spectrum, even in the thermodynamic limit. Therefore, clearly
not all initial states are able to thermalize the system. In fact, region
II is an example of these cases. However, for the Hamiltonians with $|q| > 4$%
, the kink region does not exist in the spectrum even for finite-size
condensates. Thus, we conclude that ETH holds in the strong sense for this
set of Hamiltonians.

The system for the region III does not equilibrate or show any
collapse-revival phenomena and instead oscillates, because the effective dimension saturates at a finite value whereas the Hilbert space dimension diverges in the thermodynamic limit and the main
contribution to the dynamics comes from nonthermal states which also have low
participation ratio values in the Fock basis. We explicitly showed that the
thermal and nonthermal states in the spectrum have high and low PR values with system-size scaling exponents of $\sim 0.9$ and $\sim 0.2$, respectively. In the end, thermalization seems to be linked to the
localization properties of the eigenstates. In region IV, the system
thermalizes with very small amplitude collapse and revivals either at $0$ or 
$1$. The initial state is already in almost-equilibrium with the quench
Hamiltonian, leading to almost-no nonequilibrium evolution for the system
to pursue. For the anti-ferromagnetic sudden quench map Fig.\ \ref{fig:PD2},
we always observe only the regions III and IV given that the initial state
is a ground state of an anti-ferromagnetic Hamiltonian. Finally, we note
that the region around $q_{f}\sim 0$ on both sudden quench maps is special
in terms of how the thermalization value is independent of the initial state
chosen. This behaviour is expected, because almost all of the eigenstates in the
spectrum contribute to the observable expectation value in the same amount regardless of the system size.

Interpretation of sudden quench maps as non-equilibrium phase diagrams and the transitions between regions as the dynamical phase transitions seems possible given that these dynamical transition points originate from the equilibrium quantum phase transitions of the system. We leave the question if these transitions can be related to dynamical quantum phase transitions (DQPTs) \cite{2017arXiv170907461H} as an investigation for future.

Spinor Bose-Einstein condensates are relatively more convenient to
experiment with \cite%
{PhysRevA.90.023610,PhysRevLett.102.125301,PhysRevLett.116.155301,RevModPhys.85.1191,Hoang23082016}
and numerically less costly (when SMA is applied), compared to more popular
models such as Bose-Hubbard model or Ising models. Here, we showed that
spinor BECs can also be used as a test-bench to test the ideas on the
thermalization of isolated quantum systems.

\acknowledgements
%%%%%%%%%%%%%%%%%%%%%%%%%%%%%%%%%%%%%%%%%%%%%%%%%%%%%%%%%%%
This work was supported by the ARL, the IARPA LogiQ program, and the AFOSR MURI program. C.B.D. thanks Thomas Shaw for his helpful discussions %and emotional support 
during this work.
%%%%%%%%%%%%%%%%%%%%%%%%%%%%%%%%%%%%%%%%%%%%%%%%%%%%%%%%%%%
\\
\appendix

\section{Microcanonical Window Selection}

\label{sec:appA}

The microcanonical ensemble (MC) prediction should not depend on the size of
the energy window. This constraint prevents us to calculate the MC
prediction for the cases where the kink structure exists in the spectrum and
the initial state is chosen in such a way that it overlaps with the kink.
See Figs. \ref{fig:thermalization1} and \ref{fig:thermalization2} for these
regions where the MC energy interval cannot be well-defined. This result is
consistent with the condition Eq. \ref{condition} which does not hold for
the aforementioned cases above. However the typical eigenstates of a spinor
BEC system are thermal with high PR values and therefore we can compare the
prediction of diagonal ensemble (PDE) (or the long-time average of dynamical
response) with the MC prediction for almost any initial state. 
\begin{figure}[tbp]
\centerline{\includegraphics[width=0.49%
\textwidth]{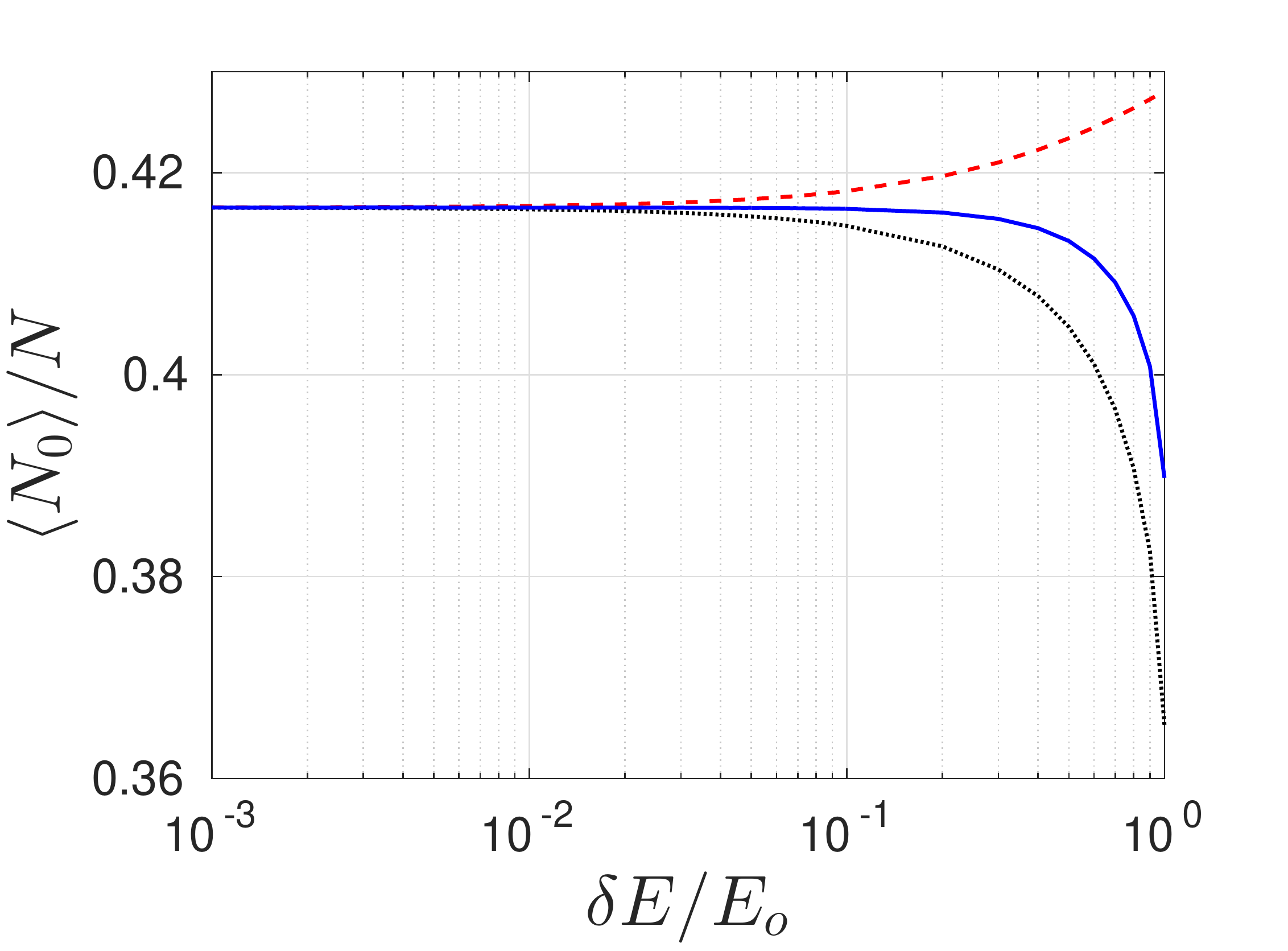}}
\caption{(Color online) The microcanonical ensemble thermal prediction with
respect to different energy intervals for $[E_o - \protect\delta E,E_o]$
(red-dashed), $[E_o - \protect\delta E,E_o+ \protect\delta E]$ (blue-solid) and $%
[E_o,E_o+ \protect\delta E]$ (black-dotted) when a sudden quench is applied from $%
q_i = -3$ to $q_f = -0.5$ for a condensate size of $N=10^4$.}
\label{fig:figAppAFig1}
\end{figure}
%(See Eq. ... for the definition of thermalization). 
For these cases, we calculate the mean energy of the system according to 
\begin{eqnarray}
E_o &=& \sum_{\alpha} |C_{\alpha}|^2 E_{\alpha},
\end{eqnarray}
where $E_{\alpha}$ is the energy associated with each eigenstate. Keeping in
mind that the energy window should be much smaller than the mean energy $%
\delta E \ll E_o$, we look for the threshold window size $\delta E_{\text{th}%
}$ that starts to affect the MC prediction. Then any $\delta E < \delta E_{%
\text{th}}$ gives a well-defined MC energy window. We also compare three
different possibilities for the window size as $[E_o - \delta E,E_o]$, $[E_o
- \delta E,E_o+ \delta E]$ and $[E_o,E_o+ \delta E]$. Fig. \ref%
{fig:figAppAFig1} shows an example of this procedure.

\section{Mapping of a Spinor Hamiltonian onto a Single Quantum-Particle Hopping Model}\label{sec:appC}

Here we show how the parameters of single quantum-particle model (Eq. \eqref{mappedHamiltonian}) depend on the sites of the lattice. Upon comparing with the spinor Hamiltonian Eq. \eqref{hamiltonian}, we observe that the Zeeman field strength $q$ modifies only the diagonal terms and hence the onsite potential terms $\eta$. Therefore, the single particle Hamiltonian family that can produce the dynamics in this paper consists of only different onsite potential configurations.
\begin{figure}[tbp]
\centering
\subfloat[]{\label{fig:appCFig1a}\includegraphics[width=0.24%
\textwidth]{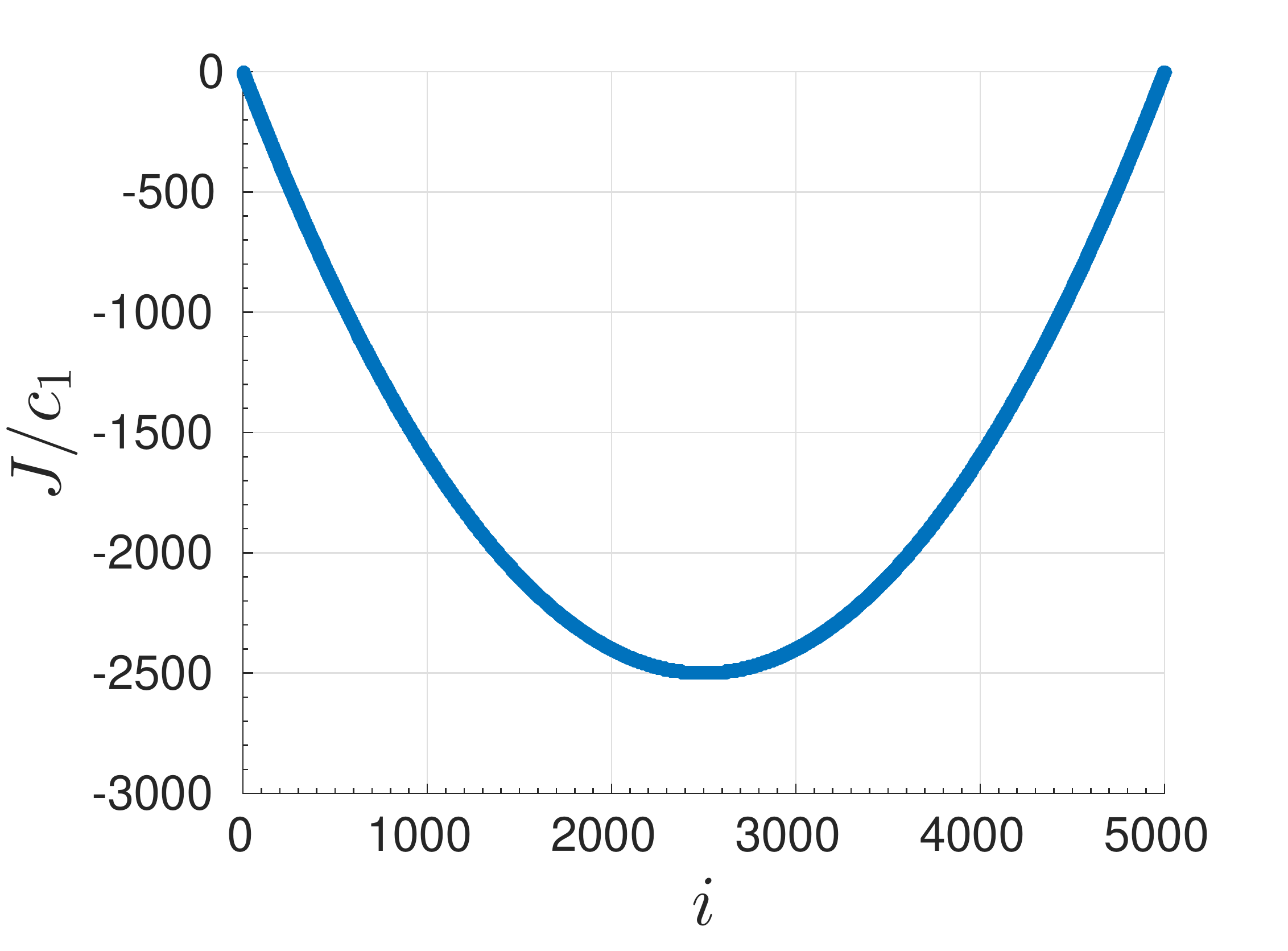}}\hfill \subfloat[]{%
\label{fig:appCFig1b}\includegraphics[width=0.24%
\textwidth]{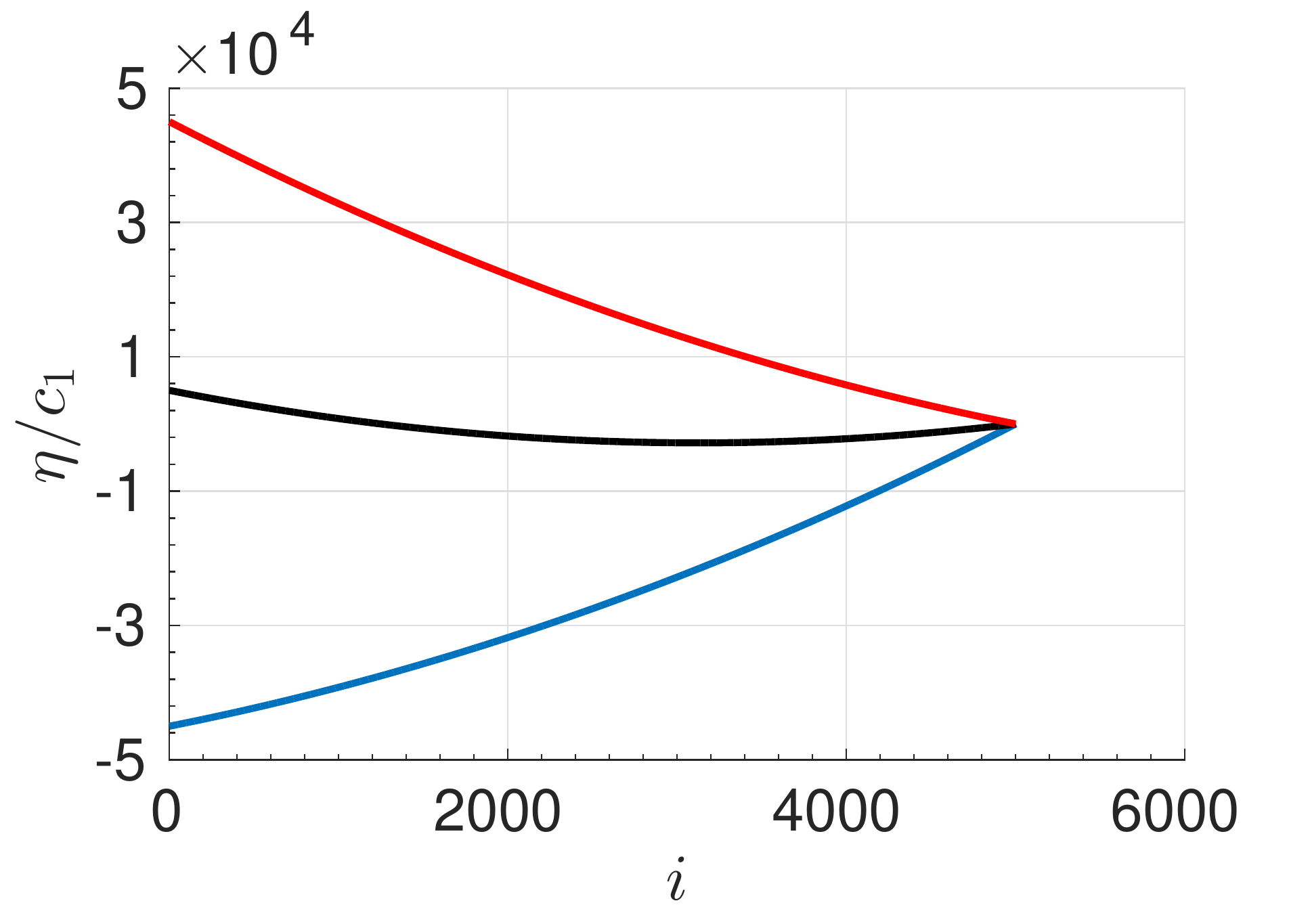}}
\caption{(Color online) (a) The hopping parameter $J$ for the mapped single particle model, (b) the onsite potential parameter $\eta$ for a condensate model with Zeeman field strength $q=4.5$ (blue-lower curve), $q=-0.5$ (black middle curve) and $q=-4.5$ (red upper curve) with respect to site position $i$ for a condensate of size $N=10^4$.}
\label{fig:appCFig1}
\end{figure}
Fig. \ref{fig:appCFig1a} shows the hopping coefficients with respect to single particle lattice positions. This functional dependence of $J$ onto the site positions is fixed for each spin-1 BEC Hamiltonian. Fig. \ref{fig:appCFig1b} shows different onsite potential configurations depending on the Zeeman field strength. The most important observation is that onsite potentials for all cases are not random, instead they are engineered potentials with respect to site positions. This property breaks the localization of single particle hopping model and hence we observe thermalization of an observable that is nonlocal for the model.

%\pagebreak 
\bibliographystyle{apsrev4-1}
%\bibliography{Bibliography}
% The references (bibliography) information are stored in the file named "Bibliography.bib"


\end{document}